\documentclass[sigconf, screen=true, review=false, printacmref=false, printccs=false, printfolios=false]{acmart}
\pdfoutput=1
\setcopyright{none}

\usepackage{xcolor}
\definecolor{thedarkblue}{RGB}{0,0,120} 
\definecolor{mydarkblue}{rgb}{0,0.08,0.45} 
\definecolor{darkblue}{rgb}{0,0.08,180}
\colorlet{TufteRed}{red!80!black}
\definecolor{theblue}{RGB}{0,0,180}
\colorlet{thered}{TufteRed}
      
\usepackage{microtype}
\usepackage{balance}

\usepackage{booktabs}
\usepackage{tabularx}

\usepackage{amsmath,amssymb,amsthm}

\newcommand{\eat}[1]{\ignorespaces}
\usepackage{comment}

\usepackage{tikz}
\usepackage{verbatim}
\usetikzlibrary{arrows}
\usetikzlibrary{shapes,snakes}
\usetikzlibrary{decorations.pathmorphing} 
\usetikzlibrary{fit}					
\usetikzlibrary{backgrounds}	

\usepackage{ragged2e}
\usepackage{multirow}
\usepackage{microtype}
\usepackage{balance}
\usepackage{setspace}

\graphicspath{{./}{./graphics/}}
\newcolumntype{H}{>{\setbox0=\hbox\bgroup}c<{\egroup}@{}}

\newcolumntype{R}[1]{>{\RaggedLeft\arraybackslash}} 
\newcolumntype{L}[1]{>{\RaggedRight\arraybackslash}} 

\newcommand{\abs}[1]{\left|#1\right|}

\newcommand{\eg}{\emph{e.g.}}
\newcommand{\ie}{\emph{i.e.}}

\newtheorem*{note}{\hspace{-1em}\textsc{Note}}
\newtheorem{property}{Property}
\newtheorem{lemma}{\hspace{-1em}\bfseries{Lemma}}
\newtheorem{Definition}{\hspace{-1em}\bfseries{Definition}}
\newtheorem{Claim}{Claim}

\AtBeginEnvironment{pmatrix}{\setlength{\arraycolsep}{2pt}}

\providecommand{\mat}[1]{\boldsymbol{\mathrm{#1}}}%
\renewcommand{\vec}[1]{\boldsymbol{\mathrm{#1}}}

\DeclareMathOperator*{\argmin}{argmin}
\DeclareMathOperator*{\argsort}{arg\,sort}

\DeclareMathOperator{\hugeE}{\mbox{\huge\raise-0.3ex\hbox{E}}}
\DeclareMathOperator{\p}{\mathbb{P}}
\DeclareMathOperator{\hugep}{\mbox{\huge\raise-0.3ex\hbox{$\p$}}}

\newcommand{\RR}{\mathbb{R}}

\providecommand{\mW}{\ensuremath{\mat{W}}}
\providecommand{\mX}{\ensuremath{\mat{X}}}
\providecommand{\mY}{\ensuremath{\mat{Y}}}

\providecommand{\vb}{\ensuremath{\vec{b}}}

\providecommand{\vd}{\ensuremath{\vec{d}}}
\providecommand{\ve}{\ensuremath{\vec{e}}}

\providecommand{\vh}{\ensuremath{\vec{h}}}

\providecommand{\vp}{\ensuremath{\vec{p}}}

\providecommand{\vs}{\ensuremath{\vec{s}}}

\providecommand{\vx}{\ensuremath{\vec{x}}}
\providecommand{\vy}{\ensuremath{\vec{y}}}

\usepackage{subfigure}
\usepackage{graphbox}
\expandafter\def\csname ver@subfig.sty\endcsname{}
\usepackage{svg}
\usepackage{nicefrac}

\usepackage{todonotes}

\DeclareMathAlphabet{\mathbcal}{OMS}{cmsy}{b}{n}

\newcommand{\confPaper}[1]{}
\newcommand{\sparseConfigVector}{$\vs_c$}
\newcommand{\denseConfigVector}{$\vd_c$}
\newcommand{\sparseVarVector}{$\vs_{x}$}
\newcommand{\denseVarVector}{$\vd_{x}$}
\newcommand{\sparseConfigVectorNoEncloMath}{\vs_c}
\newcommand{\denseConfigVectorNoEncloMath}{\vd_c}
\newcommand{\sparseVarVectorNoEncloMath}{\vs_{x}}
\newcommand{\denseVarVectorNoEncloMath}{\vd_{x}}

\usepackage{amsmath}
\usepackage{mathrsfs}
\usepackage{comment}

\usepackage{bm}
\usepackage{bbm}
\usepackage{amssymb}
\usepackage{enumitem}

\AtBeginDocument{
  \providecommand\BibTeX{{
    \normalfont B\kern-0.5em{\scshape i\kern-0.25em b}\kern-0.8em\TeX}}}

\setcopyright{none}
\acmPrice{15.00}
\acmISBN{978-1-4503-xxxx-x/18/06}
\acmConference[WWW '21]{Proceedings of the Web Conference 2021}{April 19--23, 2021}{Ljubljana, Slovenia}
\acmBooktitle{Proceedings of the Web Conference 2021 (WWW '21), April 19--23, 2021, Ljubljana, Slovenia}
\acmPrice{}
\acmDOI{10.1145/3366424.xxxxxxx}

\begin{document}
\title{\emph{ML-based Visualization Recommendation}: \\Learning to Recommend Visualizations from Data}

\settopmatter{authorsperrow=4}

\author{Xin Qian}
\affiliation{
  \institution{University of Maryland}
  \city{College Park}
  \state{MD}
  \country{USA}
}

\author{Ryan A. Rossi}
\orcid{1234-5678-9012-3456}
\affiliation{
  \institution{Adobe Research}
  \city{San Jose}
  \state{CA}
  \country{USA}
}

\author{Fan Du}
\affiliation{
  \institution{Adobe Research}
  \city{San Jose}
  \state{CA}
  \country{USA}
}

\author{Sungchul Kim}
\affiliation{
  \institution{Adobe Research}
  \city{San Jose}
  \state{CA}
  \country{USA}
}

\author{Eunyee Koh}
\affiliation{%
  \institution{Adobe Research}
  \city{San Jose}
  \state{CA}
  \country{USA}
}

\author{Sana Malik}
\affiliation{
  \institution{Adobe Research}
  \city{San Jose}
  \state{CA}
  \country{USA}
}

\author{Tak Yeon Lee}
\affiliation{
  \institution{Adobe Research}
  \city{San Jose}
  \state{CA}
  \country{USA}
}

\author{Joel Chan}
\affiliation{
  \institution{University of Maryland}
  \city{College Park}
  \state{MD}
  \country{USA}
}
\email{}

\renewcommand{\shortauthors}{X. Qian et al.}

\captionsetup[table]{skip=5pt}

\begin{abstract}
Visualization recommendation seeks to generate, score, and recommend to users useful visualizations automatically, and are fundamentally important for exploring and gaining insights into a new or existing dataset quickly. In this work, we propose the first end-to-end ML-based visualization recommendation system that takes as input a large corpus of datasets and visualizations, learns a model based on this data. Then, given a new unseen dataset from an arbitrary user, the model automatically generates visualizations for that new dataset, derive scores for the visualizations, and output a list of recommended visualizations to the user ordered by effectiveness. We also describe an evaluation framework to quantitatively evaluate visualization recommendation models learned from a large corpus of visualizations and datasets. Through quantitative experiments, a user study, and qualitative analysis, we show that our end-to-end ML-based system recommends more effective and useful visualizations compared to existing state-of-the-art rule-based systems. Finally, we observed a strong preference by the human experts in our user study towards the visualizations recommended by our ML-based system as opposed to the rule-based system (5.92 from a 7-point Likert scale compared to only 3.45).

\end{abstract}



\maketitle

\section{Introduction}
\label{sec:intro}
Nowadays, visualization has been a convenient vehicle for exploratory data analysis. 
However, due to the increasing size of real-world datasets, there are sometimes obstacles for practitioners, such as decision makers, data analysts, and researchers, to efficiently and effectively create visualizations. 
It could be overwhelming to understand an unfamiliar dataset, then select the most proper visualizations out of a myriad of valid visualization choices.
Automatic visualization recommendation systems have been developed to assist data analysts in creating visualizations. 
An end-to-end visualization recommendation system would automatically recommend a list of visualizations ordered by importance, where the visualizations uses the proper visual design to show insights about a selection of attributes\footnote{The term variable, attribute, and data column are synonyms in this work.} in the dataset. 
A successful system would greatly reduce the amount of time, cost, and effort that human spend in insight discovery process.

Previous end-to-end systems are rule-based, and leverage a small set of manually defined rules crafted by domain experts to score the generated visualizations~\cite{wongsuphasawat2015voyager,wongsuphasawat2017voyager}. As such, these rule-based systems have many issues that our proposed approach addresses. 
First, these systems often have quality issues, limiting the utility and usefulness of the recommended visualizations. 
Second, when visualizations are scored using a set of manually defined rules/heuristics, many of the visualizations receive the same exact score. This issue arises due to the way scoring is done using the rules, which often simply assigns a positive (or negative) score depending on the rule they abide by or violate, and at the end, all such scores from the small set of rules that actually apply to the visualization are combined to obtain the final score, which results in many visualizations having the same score, and thus, unable to differentiate between the visualizations receiving the same heuristic score. 
Third, adding additional rules to these systems is tedious and costly in terms of time and effort required. 
Finally, in contrast to our proposed approach that is completely automatic and data-driven, and able to adapt based on new data, or user-preferences, the existing rule-based systems are not automatic, not data-driven, and hard to iterate or improve as the preferences of users and visualizations change over time.

While there are only a few such end-to-end visualization recommendation systems, all of them are fundamentally rule-based~\cite{wongsuphasawat2015voyager,wongsuphasawat2017voyager}.
Previous work such as VizML~\cite{vizml} have used machine learning to predict the type of a chart (\eg, bar, scatter) instead of complete visualization, whereas other work such as Draco~\cite{draco} used a model to infer weights for a set of manually defined rules.
There is no such work that uses machine learning for end-to-end visualization recommendation\footnote{Note that other work used visualization recommendation more generally, however, in this work the term visualization recommendation has a very precise and formal definition to mean the recommendation of an actual visualization, not only simple design choices like chart type~\cite{vizml}, or weights for manually defined rules~\cite{draco}, etc.} in a completely automated and data-driven fashion.
This paper fills this gap by proposing the first completely automated and data-driven approach for end-to-end ML-based visualization recommendation.

In this work, we propose \textit{the first end-to-end deep learning-based visualization recommendation system} that automatically learns to generate effective and useful visualizations by leveraging previous user-generated visualizations as training.
Suppose a user selects or uploads a new dataset of interest, we can then use the learned model $\mathcal{M}$ to automatically score and recommend to the user the top-k most relevant visualizations for the users' dataset.
The approach is completely automatic, fully data-driven, flexible, and effective.
It is able to learn a general recommendation model $\mathcal{M}$ from a large corpus of datasets and their visualizations, which can then be applied for scoring and recommending visualizations for any other arbitrary dataset.

We first formalize the ML-based visualization recommendation problem and describe a general learning framework for it.
To learn a visualization recommendation model from a large corpus of training visualizations, we decompose a visualization into 
the subset of attributes selected from one of the datasets in the training corpus and a visualization configuration that describes the design choices and types of attributes required.
In particular, the proposed notion of a visualization configuration represents a data-independent abstraction where the data attributes used in the design choices are replaced by their general type.
Both of these provide us with everything required to characterize a visualization.
Next, we propose a wide-and-deep learning model for visualization recommendation based on this problem formulation that learns from the attribute selections and their visualization configurations. 
In the wide component, we learn from sparse attribute meta-features along with sparse visualization configuration features whereas in the deep component we learn from dense representations of the meta-features of the attributes and visualization configurations. 
Scores from both these components and then combined to obtain the final score of a complete visualization.

Many new evaluation issues and challenges arise when quantitatively evaluating the ranking of visualizations from an end-to-end trained ML-based visualization recommendation model.
For instance, since visualizations held-out for evaluation are from different datasets, and the space of possible visualizations to recommend differs for each dataset (as shown in Sec.~\ref{sec-problem-formulation}), then standard ranking evaluation metrics fail since the size of the visualization space changes depending on the data.
Consider two datasets, one with a large number of attributes and another with only two such attributes, then the \emph{ML-based visualization ranking problem} in the dataset with a few attributes is significantly easier than the one with a large number of attributes.
To address these challenges, 
we propose a general framework for evaluation of end-to-end ML-based visualization recommendation system.
This is the first evaluation framework for ML-based visualization recommender systems, and as such, we believe it will be useful for making further progress in developing better and more accurate end-to-end visualization recommendation systems that leverage machine learning.
The evaluation framework serves as a foundation for quantitative evaluation of future ML-based vis. rec. systems that build upon our work.

Extensive experiments evaluating the effectiveness of our approach are provided in the paper. 
Overall, the results demonstrate the effectiveness and utility of the proposed end-to-end ML-based visualization recommendation system. 
First, we conduct extensive experiments using a large-scale public corpus of datasets and user-generated visualizations. Our empirical results demonstrate the effectiveness of our approach as it is able to recover the held-out ground-truth visualizations among the large exponential space of lower quality/irrelevant visualizations. 
We also conduct a user study to investigate the quality of our wide-and-deep learning-based end-to-end visualization recommendation system compared to the state-of-the-art rule-based system. In nearly all cases, the visualizations generated and recommended by our end-to-end ML-based system are at least as good, and most often, better than those recommended by the rule-based system. 
Furthermore, we demonstrate the effectiveness of our approach through a number of case studies that clearly show a variety of important advantages of our approach compared to the existing rule-based system.

\subsection*{Main Contributions}
A summary of the main contributions of this work are as follows:
\begin{itemize}[leftmargin=*]
\item \textbf{First ML-based Visualization Recommender}:
In this work, we propose the first end-to-end \emph{ML-based visualization recommendation system} that automatically learns a model from a large set of $N$ training datasets and the corresponding $N$ sets of user-generated visualizations.
The learned model not only captures simple visual rules, but is able to learn complex high-dimensional latent characteristics behind effective user-generated visualizations from the training corpus along with the latent characteristics of the data (subset of attributes) that are associated with the visualizations.
Given a new (unseen) dataset of interest, our learned model can then generate, score, and automatically recommend the top most insightful and effective visualizations for this new dataset.\footnote{Note each visualization uses a subset of the attributes from the dataset, and some attributes may never be used.}

\item \textbf{Problem Formulation}: We carefully formalize the problem of learning a visualization recommendation model from training data consisting of $N$ datasets and $N$ sets of visualizations.\footnote{A single dataset is associated with a set of visualizations from that dataset.}
To the best of our knowledge, this is the first formal presentation of the ML-based vis. rec. problem.

\item \textbf{Evaluation Framework}: 
We describe a general framework for evaluation of end-to-end ML-based visualization recommendation systems using a held-out set of known ground-truth visualizations from a set of new held-out datasets that were not used to train the model.
The evaluation framework serves as a foundation for quantitative evaluation of future ML-based vis. rec. systems that build upon our work.

\item \textbf{Effectiveness}:
We demonstrate the effectiveness of our approach through a comprehensive set of experiments including quantitative evaluation of the visualization ranking (Sec.~\ref{sec:exp-quant-results}), user study comparing our ML-based system to a rule-based system (Sec.~\ref{sec:exp-user-study}), and a qualitative case study (Sec.~\ref{sec:exp-case-study}). 
Overall, the results demonstrate the effectiveness and utility of the proposed end-to-end ML-based visualization recommendation system. 
\end{itemize}

\section{Related Work} \label{sec:related-work}
Related work can be categorized as follows:
(i) rule-based systems that recommend entire visualizations, and 
(ii) approaches designed for simpler, but fundamentally different sub-tasks
such as predicting the chart type of a visualization.
Despite that they do not solve the end-to-end visualization recommendation problem, we include them since some of them use machine learning to solve a fundamentally different problem.

Systems that recommend entire visualizations to users have existed since 1980s~\cite{dorisjang}.
Early systems include Automatic Presentation Tool (APT)~\cite{mackinlay1986automating} that generates graphical presentations from data, SAGE~\cite{roth1994interactive} that uses a search algorithm to select and compose graphics from data, and ShowMe~\cite{mackinlay2007show}.  
These systems can be attributed as rule-based solutions.
For example, Automatic Presentation Tool (APT)~\cite{mackinlay1986automating} uses logical statements of visualization design knowledge that come from human perceptual experiments. 
SAGE~\cite{roth1994interactive} generates visualizations based on partial specifications. 
ShowMe~\cite{mackinlay2007show} offers users with a set of defaults to filter valid visualization types based on characteristics of selected data. 
Mixed-initiative systems such as Voyager~\cite{vartak2017towards, wongsuphasawat2017voyager, wongsuphasawat2015voyager}, VizDeck~\cite{perry2013vizdeck}, and DIVE~\cite{hu2018dive} incorporate visual encoding rules to assist user data exploration data exploration. 
Rule-based systems have many limitations that our work addresses.
For instance, rule-based recommendation systems rely entirely on a large set of manually defined rules from domain experts, which are costly in terms of the manual labor required, and may miss many important rules that would provide users with significantly more effective visualizations for their dataset of interest.
Such approaches are clearly not data-driven and difficult to adapt as one would need to routinely incorporate new rules in a manual fashion, which is costly in terms of the time and effort required by domain experts to maintain such systems.
Further, rules for such systems need to be manually defined with respect to the domain of interest.
For instance, visualizations for data scientists, or scientific domains are likely different from visualizations that journalists prefer or those that would work well for elderly populations.
Therefore, new rule sets would likely be required to effectively recommend visualizations for each group.
These systems also require tailored experiments with human users to validate the manually defined rules.
In comparison, our work learns a model $\mathcal{M}$ to recommend entire visualizations directly from a large corpus of training data, in a fully automatic data-driven fashion. 
Furthermore, we also propose an evaluation framework to validate the effectiveness of ML-based visualization recommendation models.

On the other hand, there are systems that tackle sub-tasks in visualization recommendation. 
Each of those systems has a distinct focus in some end goals such as improving expressiveness, improving perceptual effectiveness, matching user task types, etc. 
The sub-tasks can generally be divided two categories~\cite{dorisjang, wongsuphasawat2016towards}: whether the solution focuses on recommending data (\textit{what data to visualize}), such as Discovery-driven Data Cubes~\cite{sarawagi1998discovery}, Scagnostics~\cite{wilkinson2005graph}, AutoVis~\cite{wills2010autovis}, and MuVE~\cite{ehsan2016muve}) or recommending encoding (\textit{how to design and visually encode the data}), such as APT~\cite{mackinlay1986automating}, ShowMe~\cite{mackinlay2007show}, and Draco--learn~\cite{draco}). 
While some of those are ML-based, none recommends entire visualizations, and thus does not solve the visualization recommendation problem that lies at the heart of our work.
For example, VizML~\cite{vizml} 
used machine learning to predict the type of a chart (e.g., bar,scatter, etc.) instead of complete visualization.
Another work Draco~\cite{draco} used a model to infer weights for a set of manually defined rules.
VisPilot~\cite{lee2019avoiding} recommended different drill-down data subsets from datasets. 
Instead of solving simple sub tasks such as predicting the chart type of a visualization, 
we focus on the \emph{end-to-end visualization recommendation task} where 
the goal is to \textit{automatically recommend users the top-k most effective visualizations as the output, given an input dataset from the user}.

This paper fills the gap by proposing the first end-to-end ML-based visualization recommendation approach that is completely automatic and data-driven.
It tackles both choosing data from datasets, and recommending encoding for selected data, therefore achieving the goal of recommending complete visualizations from arbitrary datasets using an automatically learned model $\mathcal{M}$ from a large corpus of training data.

\section{ML-based Problem Formulation}\label{sec-problem-formulation}
In this section, we formally introduce the ML-based visualization recommendation problem, and present a generic learning framework for it, which includes two key parts:
\begin{itemize} 
    \item \textbf{Model Training (Sec.~\ref{sec:problem-model-training}):} 
    Given a training visualization corpus $\mathbcal{D} = \{\mX_i, \mathbb{V}_{i}\}_{i=1}^N$ consisting of $N$ datasets
    $\{\mX_i\}_{i=1}^N$ and the corresponding $N$ sets of
    visualizations $\{\mathbb{V}_i\}_{i=1}^N$,\footnote{$\mathbb{V}_i$ is the set of visualizations associated with the $i$th dataset $\mX_i$.}
    we first 
    learn a model $\mathcal{M}$ from the training corpus $\mathbcal{D}$ that best captures and scores effective visualizations highly and assigns low scores to  bad/ineffective visualizations.\footnote{The learned model $\mathcal{M}$ not only captures simple visual rules, but is able to learn complex high-dimensional latent characteristics behind effective user-generated visualizations from the training corpus along with the latent characteristics of the data (subset of attributes) that are associated with the visualizations.}
    
    \item \textbf{Recommending Visualizations (Sec.~\ref{sec:problem-recommending-vis}):} 
    Given a new (unseen) dataset $\mX_{\rm test} \not\in \mathbcal{D}$ of interest, our learned visualization recommendation model $\mathbcal{M}$ is used to generate, score, and automatically recommend the top most insightful and effective visualizations for this new dataset.\footnote{Note each visualization uses a subset of the attributes from the dataset, and some attributes may never be used.}
\end{itemize}
Notice that the fundamental difference between the rule-based visualization recommendation problem and our proposed ML-based visualization recommendation problem is that ML-based models are automatically learned from data whereas rule-based approaches are manually defined (and are not true models).

The visualization recommendation training data $\mathbcal{D}=\{\mX_i, \mathbb{V}_i\}_{i=1}^N$ can be general, as it can consist of a set of datasets and a set of relevant visualizations from each dataset collected from a variety of different sources.\footnote{Hence, each dataset has a corresponding set of visualizations that use a subset of attributes from the dataset.}
For instance, the corpus may consist of datasets and visualizations collected from websites (\eg, by crawling the web) or from a visual analytic platform such as Tableau and Power BI where users upload datasets and created corresponding visualizations. 
Depending on the corpus, the definition of a visualization to be effective is also flexible and that reflects how users from that corpus source perceive as effective visualizations.
For example, a visualization in a data journalism website emphasizes attractiveness while a visualization in scientific papers need to be straightforward and scientifically meaningful.
We use the corpus to train the ML-based vis. rec. model.

Each visualization uses a subset of attributes from a dataset $\mX_i$, which we call the subset as the \textit{attribute combination}.
We now define the space of attribute combinations $\mathbcal{X}_i = \{\mX_i^{1},\!\ldots, \mX_i^{(k)},\!\ldots\}$ for an arbitrary dataset $\mX_i$, which can be either a training dataset $\mX_i \in \mathbcal{D}$ or a new test dataset $\mX_{\rm test} \leftarrow \mX_i \not\in \mathbcal{D}$.

\begin{Definition}[Space of Attribute Combinations]\label{def:attribute-combination-space}
Given an arbitrary dataset matrix $\mX_i$, let $\mathbcal{X}_i$ denote the space of attribute combinations of $\mX_i$ defined as
\begin{align}
    \Sigma : \mX_i \to \mathbcal{X}_i, \quad \text{s.t.} \label{eq:attr-comb-generator-func} \\
    \mathbcal{X}_i = \{\mX_i^{(1)},\ldots,\mX_i^{(k)},\ldots\}, \label{eq:attr-combinations-of-a-dataset}
\end{align}\noindent
where $\Sigma$ is an attribute combination generation function 
and every $\mX_i^{(k)} \in \mathbcal{X}_i$ is a different subset (combination) of attributes from $\mX_i$, and thus $\mX_i^{(k)}$ may consist of one, two, or more attributes from $\mX_i$.
\end{Definition}\noindent
When the dataset $\mX_i$ is a new test dataset $\mX_i \not\in \mathbcal{D}$, we use $\mX_{\rm test}$ to denote the new unseen dataset and the space of attribute combinations from the new test dataset is $\mathbcal{X}_{\rm test}$.

Let $|\mX_i\!|$ and $|\mX_j\!|$ denote the number of attributes (columns) of two arbitrary datasets $|\mX_i\!|$ and $|\mX_j\!|$, respectively.
\begin{property}\label{prop:comparing-var-comb-spaces}
If $|\mX_i\!|>|\mX_j\!|$, then $|\mathbcal{X}_i|>|\mathbcal{X}_j|$.
\end{property}\noindent
The proof of Property~\ref{prop:comparing-var-comb-spaces} is straightforward, but Property~\ref{prop:comparing-var-comb-spaces} will be important later when characterizing the space of possible visualizations from a given dataset.

\begin{Definition}[Space of Visualization Configurations] \label{def:space-vis-configs}
Let $\mathbcal{C}$ denote the space of all visualization configurations such that 
a visualization configuration $\mathcal{C}_{ik} \in \mathbcal{C}$ defines
an abstraction of a visualization where for each visual design choice (x, y, marker-type, color, size, etc.) that maps to an attribute in $\mX_i$, we replace it with its type.
Therefore visualization configurations are essentially visualizations without any attribute (data).
\end{Definition}

\begin{property}\label{prop:visualization-configuration-independent}
Every visualization configuration $\mathcal{C}_{ik} \in \mathbcal{C}$ is independent of any data matrix $\mX$ (by Definition~\ref{def:space-vis-configs}).
\end{property}
The above implies that $\mathcal{C}_{ik} \in \mathbcal{C}$ can potentially arise from any arbitrary dataset and is therefore not tied to any specific dataset since visualization configurations are general abstractions where the data bindings have been replaced with their general type (\eg, if x/y in some visualization mapped to an attribute in $\mX_i$, then it is replaced by the type of that attribute, that is, ordinal, quantitative, categorical, etc.
\begin{figure*}[h]
\begin{center}
\includegraphics[width=0.85\linewidth]{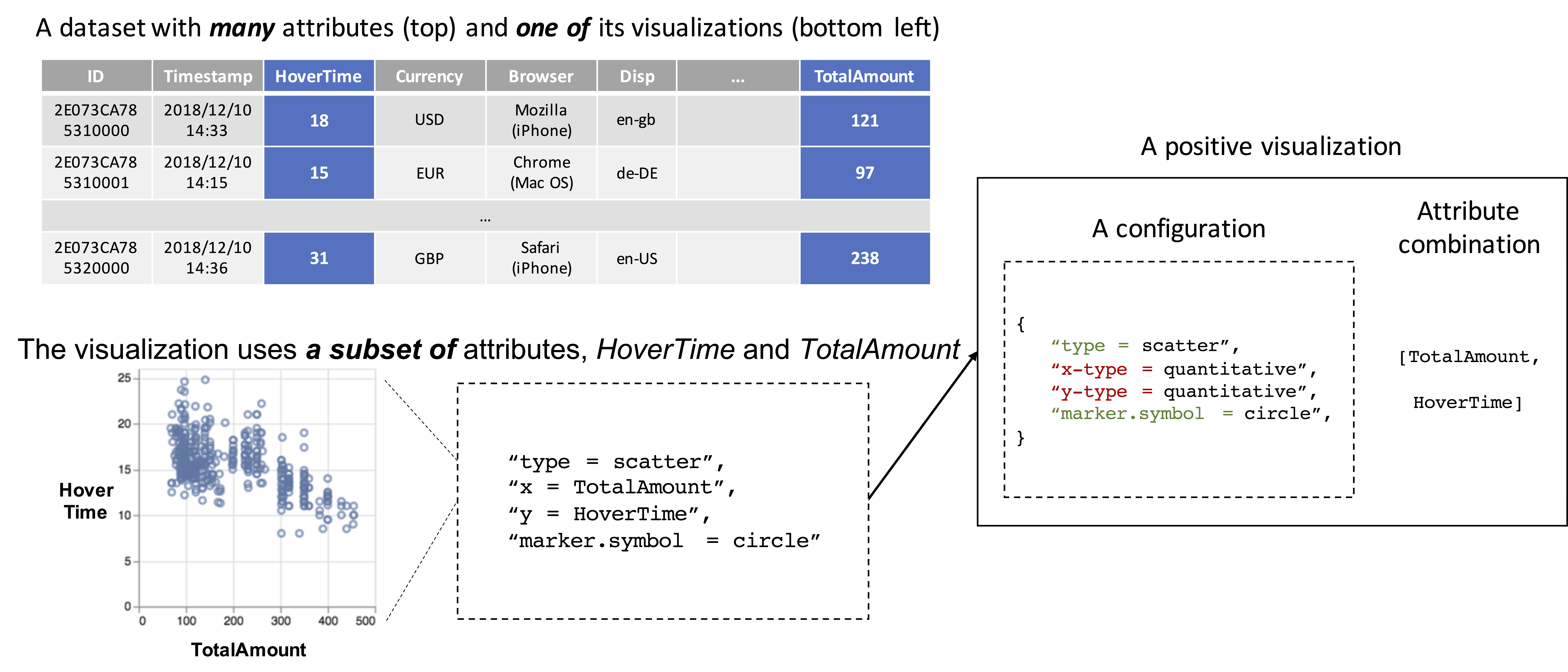}
\end{center}
\caption{The process of extracting positive training visualizations.
The left figure shows a dataset from the corpus. 
The dataset has a set of visualizations.
One visualization uses a subset of attributes from the dataset.
The right figure is an extracted positive visualization that separates the visualization into a configuration and attribute selection.
The visualization will be used for training the visualization recommendation model. 
}
\label{fig-plotly-schema-example}
\end{figure*}

A visualization configuration \emph{and} the attributes selected\footnote{Selected attributes is the same as the combination of attributes defined in Eq.~\ref{eq:attr-combinations-of-a-dataset}.}%
is everything necessary to generate a visualization.
See Figure~\ref{fig-plotly-schema-example} for an example.
The size of the space of visualization configurations 
is large since visualization configurations come from all possible combinations of design choices and their values such as,
\begin{itemize}
\item \textbf{mark/chart:} bar, scatter, ...
\item \textbf{x-type:} quantitative, nominal, ordinal, temporal, ..., none
\item \textbf{y-type:} quantitative, nominal, ordinal, temporal, ..., none
\item \textbf{color:} red, ..., quantitative, nominal, ordinal, temporal, ...
\item \textbf{size:} 1pt, 2pt, ..., quantitative, nominal, ordinal, temporal, ...
\item \textbf{x-aggregate:} sum, mean, bin, ..., none
\item \textbf{y-aggregate:} sum, mean, bin, ..., none
\item \textbf{...} 
\end{itemize}

Recall that $\Sigma$ is an attribute combination generation function defined as $\Sigma : \mX_i \to \mathbcal{X}_i$
where $\mathbcal{X}_i$ is the space of all combinations of attributes from dataset $\mX_i$, \ie, all subsets of one or more attributes from $\mX_i$.
For instance, suppose we have a dataset with three attributes $\mX = [\, \vx_{1} \; \vx_{2} \; \vx_{3} \,]$, then $\Sigma(\mX) = \mathbcal{X}$ is:
\begin{equation}\label{eq:attr-comb-generator-example}
\Sigma(\mX) = 
\big\{
\underbrace{\vx_{1}, \vx_{2}, \vx_{3}}_{\Sigma_1(\mX)}, 
\underbrace{%
[\, \vx_{1} \, \vx_{2} \,], 
[\, \vx_{1} \, \vx_{3} \,],
[\, \vx_{2} \, \vx_{3} \,]
}_{\Sigma_2(\mX)},
\underbrace{%
[\, \vx_{1} \, \vx_{2} \, \vx_{3} \,]
}_{\Sigma_3(\mX)}
\big\}
\end{equation}

\begin{Definition}[Space of Visualizations of $\mX_i$] \label{def:space-of-visualizations}
Given an arbitrary dataset matrix $\mX_i$, we define $\mathbb{V}^{\star}_i$ as the space of all possible visualizations that can be generated from $\mX_i$.
More formally, the space of visualizations $\mathbb{V}^{\star}_i$ is defined with respect to a dataset $\mX_i$ and the space of visualization configurations $\mathbcal{C}$, 
\begin{align} 
    &\Sigma(\mX_i) = \mathbcal{X}_i = \{\mX_i^{1},\!\ldots, \mX_i^{(k)},\!\ldots\}\\
    &\xi : \mathbcal{X}_i \times \mathbcal{C} \to \mathbb{V}_i^{\star} \label{eq:vis-generator}
\end{align}\noindent
where $\mathbcal{X}_i = \{\mX_i^{1},\!\ldots, \mX_i^{(k)},\!\ldots\}$ is the set of all possible attribute/attribute combinations of $\mX_i$ (Def.~\ref{def:attribute-combination-space}).
More succinctly, $\xi : \Sigma(\mX_i) \times \mathbcal{C} \to \mathbb{V}_i^{\star}$, and therefore  $\xi(\Sigma(\mX_i),\mathbcal{C}) = \mathbb{V}_i^{\star}$.
\end{Definition}\noindent
In other words, given 
an attribute combination $\mX_i^{(k)} \in \mathbcal{X}_i$ consisting of a subset of attributes from dataset $\mX_i$ and a visualization configuration $\mathcal{C} \in \mathbcal{C}$, then 
$\xi(\mX_i^{(k)}, \mathcal{C})$
is the corresponding visualization.
Define $\mX \not= \mY \implies \forall i,j\,\, \vx_i \not= \vy_j$.
\begin{lemma}\label{lem:vis-space-data-dependence}
$\forall\mX_i$, $\mX_j$ s.t. $\mX_i \!\not=\! \mX_j$, then
$\xi(\Sigma(\mX_i),\mathbcal{C}) \cap \xi(\Sigma(\mX_j),\mathbcal{C}) = \emptyset$.
\end{lemma}
This is straightforward to see 
and implies that 
when $\mathbcal{C}$ is fixed, 
the space of visualizations is entirely dependent on the dataset, and for any two datasets $\mX_i$ and $\mX_j$ without any shared attributes/overlap $\mX_i \not= \mX_j$, then the set of possible visualizations that can be generated from either dataset are entirely disjoint from one another, that is $\mathbb{V}_{i} = \xi(\Sigma(\mX_i),\mathbcal{C})$ and $\mathbb{V}_{j} = \xi(\Sigma(\mX_j),\mathbcal{C})$ where $\mathbb{V}_{i} \cap \mathbb{V}_{j}=\emptyset$. 
Hence, $|\mathbb{V}_{i} \cap \mathbb{V}_{j}|=0$ and $\mathbb{V}_{i} \cup \mathbb{V}_{j}=|\mathbb{V}_{i}| + |\mathbb{V}_{j}|$.
If $|\mX_i| > |\mX_j|$,  then $|\xi(\Sigma(\mX_i),\mathbcal{C})| > |\xi(\Sigma(\mX_j),\mathbcal{C})|$.

\begin{Definition}[Positive Visualizations of $\mX_i$]
Given an arbitrary dataset matrix $\mX_i$, we define $\mathbb{V}_i$ as the set of positive visualizations (user-generated, observed) from dataset $\mX_i$.
Therefore, 
\begin{equation}
    \mathbb{V} = \bigcup_{i=1}^N \mathbb{V}_i \quad \text{ and } \quad |\mathbb{V}| \geq N
\end{equation}
\end{Definition}

\begin{Definition}[Negative Visualizations of $\mX_i$] \label{def:negative-vis-of-a-dataset}
Let $\mathbb{V}_i^{\star}$ denote the space of all visualizations that arise from the $i$th dataset $\mX_i$ such that the user-generated (positive) visualizations $\mathbb{V}_i$ satisfies $\mathbb{V}_i \subseteq \mathbb{V}_i^{\star}$, then the \emph{space of negative visualizations for dataset $\mX_i$} is 
\begin{equation} \label{eq:negative-vis-space-of-Xi}
\mathbb{V}_i^{-} = \mathbb{V}_i^{\star} \setminus \mathbb{V}_i 
\end{equation}\noindent
This follows from $\mathbb{V}_i^{-} \cup \mathbb{V}_i = \mathbb{V}_i^{\star}$.
\end{Definition}

\begin{note}
The space of negative visualizations between different datasets is also obviously completely disjoint.
\end{note}

Given $\mathbb{V}_i^{\star}$ as the space of all visualizations of the $i$th dataset $\mX_i$ where it consists of positive and negative visualizations, denoted as $\mathbb{V}_i^{-} \cup \mathbb{V}_i = \mathbb{V}_i^{\star}$, we define $Y_{ik}$ as the ground-truth label of a visualization $\mathcal{V}_{ik} \in \mathbb{V}_i^{\star}$, such that
\begin{equation}
Y_{ik} = \begin{cases}
            1 & \text{if \quad $\mathcal{V}_{ik} \in \mathbb{V}_i$} \\
            0 & \text{otherwise}
        \end{cases}
\end{equation}

\begin{Definition}[Sampling Negative Visualizations of $\mX_i$]\label{def:negative-vis-sampling}
Given dataset $\mX_i$, we sample negative visualizations from $\mathbb{V}_i^{-} = \mathbb{V}_i^{\star} \setminus \mathbb{V}_i$ (Def.~\ref{def:negative-vis-of-a-dataset}) as follows:
\begin{align}\label{eq:negative-vis-sampling-uniform}
   &k \sim \text{UniformDiscrete}\{1,2,\ldots,|\mathbb{V}_i^{-}|\},\quad \text{for } j=1,2,\ldots\\
    &\widehat{\mathbb{V}}_{i}^{-} = \widehat{\mathbb{V}}_{i}^{-} \cup \mathcal{V}_{ik}^{-}
\end{align}\noindent
where $\widehat{\mathbb{V}}_{i}^{-} \subseteq \mathbb{V}_{i}^{-}$.
Hence, $\mathcal{V}_{ij}^{-}$ denotes the $j$th negative visualization for dataset $\mX_i$ sampled from $\mathcal{V}_{ij}^{-} \in \mathbb{V}_{i}^{-}$.
\end{Definition}
The negative visualization space is large and therefore sampling of this vast space is required to ensure fast and computationally tractable inference.
In Eq.~\ref{eq:negative-vis-sampling-uniform}, we sample the negative visualization space of dataset $\mX_i$ uniformly at random with replacement.
Recall that any form of estimation is difficult since the size of the space of visualizations $\mathbb{V}_i^{\star}$, including positive visualizations $\mathbb{V}_i$ and negative visualizations $\mathbb{V}_i^{-}$ depends entirely on the number of attributes/attributes in the dataset $\mX_i$ (and their types, such as real-valued, ordinal, categorical, etc.) used in their generation, and thus the size of the different visualization spaces varies based on it.
Sampling negative visualizations is important for both training and testing.

\subsection{Learning Vis. Rec. Model} \label{sec:problem-model-training}
Now we formulate the problem of training a visualization recommendation model $\mathbcal{M}$ from a large training corpus of datasets and sets of visualizations associated to each dataset.

\begin{Definition}[Learning Vis. Recommendation Model] \label{def:ML-based-vis-rec-model-learning-training}
Let $\mathbcal{D} = \{\mX_i,\mathbb{V}_i\}_{i=1}^{N}$ denote the training set consisting of datasets $\{\mX_i\}_{i=1}^{N}$ and the corresponding $N$ sets of visualizations $\{\mathbb{V}_i\}_{i=1}^{N}$ for the $N$ datasets.
Given the set of training datasets and relevant visualizations $\mathbcal{D} = \{\mX_i,\mathbb{V}_i\}_{i=1}^{N}$, the goal is to learn a visualization recommendation model $\mathcal{M}$ by optimizing the following general objective function,

\begin{align}\label{eq:vis-rec-model-learning}
    \argmin_{\mathcal{M}} 
    \sum_{i=1}^{N} 
    \sum_{(\mX_i^{(k)}\!,\mathcal{C}_{ik}) \in \mathbb{V}_{i}^{-} \cup \mathbb{V}_{i}}
        \mathcal{L}\Big(
        Y_{ik} \,\big|\, 
        \Psi(\mX_i^{(k)}), f(\mathcal{C}_{ik}), \mathcal{M}
        \Big)
\end{align}\noindent
where $\mathcal{L}$ is the loss function,
$Y_{ik}=\{0,1\}$ is the ground-truth label of the $k$th visualization $\mathcal{V}_{ik} = (\mX_i^{(k)}\!,\mathcal{C}_{ik}) \in \mathbb{V}_{i}^{-} \cup \mathbb{V}_{i}$
for dataset $\mX_i$.
Further, $\mX_i^{(k)} \subseteq \mX_i$ is the combination of attributes used in the visualization.
In Eq.~\ref{eq:vis-rec-model-learning}, $\Psi$ and $f$ are general functions over the attribute combination $\mX_i^{(k)} \subseteq \mX_i$ and the visualization configuration $\mathcal{C}_{ik}$ of the visualization $\mathcal{V}_{ik} = (\mX_i^{(k)}\!,\mathcal{C}_{ik}) \in \mathbb{V}_{i}^{-} \cup \mathbb{V}_{i}$, respectively.\footnote{Note $\Psi$ and $f$ can also be learned along with the model $\mathcal{M}$ or learned/defined prior to learning the model $\mathcal{M}$.}
\end{Definition}\noindent
For computational tractability, we replace $\mathbb{V}_{i}^{-}$ in Eq.~\ref{eq:vis-rec-model-learning} with the set $\widehat{\mathbb{V}}_{i}^{-}$ of sampled negative visualizations for the $i$th dataset matrix $\mX_i$.
As an aside, we provide a general formulation of the training of the model $\mathcal{M}$ in Definition~\ref{def:ML-based-vis-rec-model-learning-training}.
Intuitively, the learned model $\mathcal{M}$ from Eq.~\ref{eq:vis-rec-model-learning} can then be used to score the effectiveness of any arbitrary visualization.
Most importantly, it even enables us to score visualizations generated from entirely new datasets not used for training $\mathcal{M}$, \ie, a dataset $\mX_{\rm test}$ outside the training corpus $\mX_{\rm test} \not\in \mathbcal{D}$.
\begin{equation}\label{eq:model-scoring}
    \mathcal{M} : \mathbcal{X}_{\rm test} \times \mathbcal{C} \to \RR
\end{equation}\noindent
Hence, given an arbitrary visualization, $\mathcal{M}$ outputs a score describing the effectiveness or importance of the visualization.
The ML-based model learning formulation for visualization recommendation shown in Eq.~\ref{eq:vis-rec-model-learning} can naturally be used to recover many different types of visualization recommendation models.

\begin{Definition}[Meta-Feature Function] \label{def:meta-feature-function}
Let $\Psi$ denote the meta-feature learning function that maps an attribute $\vx$ of any dimensionality (from any dataset $\mX$) to a shared $K$-dimensional meta-feature space that captures the important characteristics of $\vx$.
More formally, 
\begin{equation}\label{eq:meta-feature-single-var}
    \Psi : \vx \to \RR^{K}
\end{equation}\noindent
where $\vx$ can be of an arbitrary attribute type (\eg, real-valued, integral, nominal, ordinal, etc) and size, \eg, two attributes $\vx \in \mX$ and $\vy \in \mY$ from two different datasets are almost surely of different dimensionality (\# rows). 
Further, given $M$ attributes of a dataset $\mX = \{\vx_1,\vx_2,\ldots,\vx_{M}\}$, then
\begin{equation}\label{eq:meta-feature-all-var}
    \Psi : \mX \to \RR^{K \times M}
\end{equation}\noindent
Hence, from Eq.~\ref{eq:meta-feature-single-var} $\Psi(\vx) \in \RR^{K}$ and $\Psi(\mX) \in \RR^{K \times M}$.
\end{Definition}

\subsection{Recommending Visualizations via Model} \label{sec:problem-recommending-vis}
Once we have learned the visualization recommendation model $\mathcal{M}$ (Eq.~\ref{eq:vis-rec-model-learning}) using the training visualization corpus $\mathbcal{D}$, then we can use $\mathcal{M}$ to score and recommend a list of the top most important and insightful visualizations generated from an arbitrary new dataset $\mX_{\rm test} \not \in \mathbcal{D}$.

\begin{Definition}[ml-based Visualization Recommendation] \label{def:ML-based-vis-rec}
Let $\mathcal{M}$ be the trained visualization recommender model from Def.~\ref{def:ML-based-vis-rec-model-learning-training}.
Given $\mathcal{M}$ along with a new (unseen) dataset $\mX_{\rm test} \not\in \big\{\mX_i\big\}_{i=1}^{N}$ of interest, then 
\begin{equation}\label{eq:model-scoring}
    \mathcal{M} : \mathbcal{X}_{\rm test} \times \mathbcal{C} \to \RR
\end{equation}\noindent
where $\mathbcal{X}_{\rm test}=\{\ldots,\mX^{(k)}_{\rm test} ,\ldots\}$ is the space of attribute combinations from $\mX_{\rm test}$ and $\mathbcal{C}$ is the space of visualization configuration.
Given the set of generated visualizations 
$\mathbb{V}_{\rm test} = \{\mathcal{V}_1, \mathcal{V}_2, \ldots, \mathcal{V}_{Q}\}$,
we derive a ranking of the visualizations $\mathbb{V}_{\rm test}$ from $\mX_{\rm test}$ as follows:
\begin{equation} \label{eq:ranking-vis}
\rho\big(\{\mathcal{V}_{1},\mathcal{V}_{2},\ldots,V_{\mathcal{Q}}\}\big)\, =\; \argsort_{\mathcal{V}_{t} \in \mathbb{V}_{\rm test}} \; \mathcal{M}(\mathcal{V}_{t})
\end{equation}\noindent
where $Q = |\mathbb{V}_{\rm test}|$.
Hence, given an arbitrary visualization, $\mathcal{M}$ outputs a score describing the effectiveness or importance of the visualization.
\end{Definition}

Informally, given a \emph{new dataset} $\mX_{\rm test}$
to recommend visualizations for via the trained model $\mathcal{M}$ (Eq.~\ref{eq:vis-rec-model-learning}), then 
$\mathcal{M}(\xi(\Sigma(\mX_{\rm test}), \mathbcal{C}))$ where $\mathbcal{C}$ is the space of relevant visualization configurations.
Notice that $\mathcal{M}(\mathbb{V}_{\rm test}) = \mathcal{M}(\xi(\Sigma(\mX_{\rm test}), \mathbcal{C}))$.
For tractability, we replace the set of possible visualization configurations $\mathbcal{C}$ with the set of relevant configurations $\mathbcal{C}_r = \mathcal{R}(\mathbcal{C})$ where $\mathcal{R}$ is a function consisting of visual rules that enables us to discard configurations that are invalid with respect to the manually defined rules. The list of rules from Voyager and other rule-based systems can read from a file similar to stopwords in information retrieval.
Hence, $\mathbcal{C}_r \subseteq \mathbcal{C}$.

Given a new dataset of interest, the space of visualizations to search over is completely different from the space of visualizations that arises from any other (non-identical) dataset.
More formally, 
let $\mathbb{V}_i^{\star}$ and $\mathbb{V}_j^{\star}$ denote the space of all possible visualizations that arise from $\mX_i$ and $\mX_j$ held-out datasets, then $\forall r,s$, $\mathcal{V}_{r} \in \mathbb{V}_{i}^{\star} \not= \mathcal{V}_{s} \in \mathbb{V}_{j}^{\star}$ holds.
Further, this obviously holds $\forall i,j \in [T]$ as well.
Clearly, the above holds, since a visualization consists of a subset of attributes (data) and design choices.
The above demonstrates the difficulty of the visualization recommendation learning problem, in the sense that, the model must recommend relevant visualizations from a space of visualizations never seen by the learning algorithm.
Moreover, we can even show a weaker property regarding the cardinality of the space of visualizations that arise from different held-out datasets,
\begin{Claim}
Let $\mathbb{V}_i^{\star}$ and $\mathbb{V}_j^{\star}$ denote the space of all possible visualizations that arise from $\mX_i$ and $\mX_j$ held-out datasets, then with high probability $|\mathbb{V}_i^{\star}| \not= |\mathbb{V}_j^{\star}|$ almost surely holds $\forall i,j \in [T]$.
\end{Claim}

\section{Wide \& Deep Visualization Recommendation} 
\label{sec-approach}
Following the general ML-based visualization recommendation formulation in Section~\ref{sec-problem-formulation}, we now describe our proposed wide-and-deep visualization recommendation approach.
Table~\ref{tab-summary-of-notation-approach} provides a summary of the key notation.

\begin{figure}[b!]
\begin{center}
\includegraphics[width=1.0\linewidth]{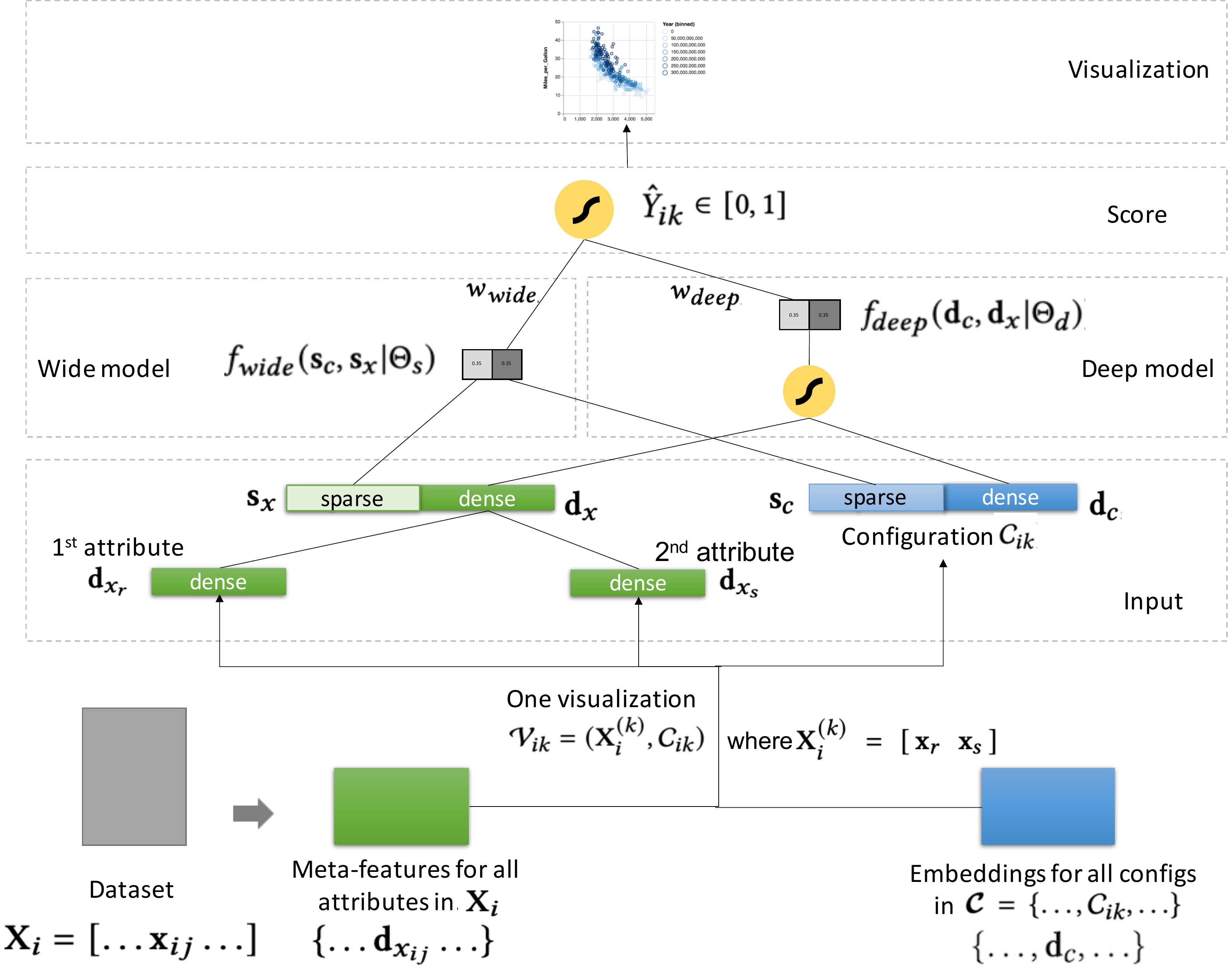}
\end{center}
\caption{
For an arbitrary dataset $\mX_i$ (either a new unseen dataset or one from the training corpus), we generate the space of visualizations $\mathbb{V}^{\star}_i$ for $\mX_i$.
The visualizations then feed our wide-and-deep network model $\mathcal{M}$ one-by-one. 
For each visualization, the network takes as input the attribute combination $\mX_i^{(k)}$ and a configuration $\mathcal{C}_{ik}$, and outputs a score $\hat{Y}_{ik}$ as the predicted effectiveness of this visualization. 
}
\label{fig-joint-modeling}
\end{figure}

\subsection{Wide-and-Deep Network Overview}
\label{sec-network-overview}
We now give a brief overview of the wide-and-deep learning-based visualization recommendation approach.
Figure~\ref{fig-joint-modeling} shows the wide-and-deep network architecture.

\begin{itemize}[leftmargin=*]
    \item \textbf{Encoding Visualizations and Their Attributes (Sec.~\ref{sec-encode-input}):} 
    The network first encodes the visualization $\mathcal{V}_{ik}$ from one arbitrary dataset $\mX_{i} \in \mathbcal{D}$ by its attribute combination $\mX_i^{(k)}$ and the visualization configuration $\mathcal{C}_{ik}$ into dense and sparse features (Section~\ref{sec-encode-input}), denoted as $\vd_x$, $\vd_c$, $\vs_x$ and $\vs_c$. 
    
    \item \textbf{\emph{Wide} Vis. Rec. Model (Sec.~\ref{sec-wide-model}):} 
    The \emph{wide} model takes as input the sparse features $\vs_x$ and $\vs_c$ of $\mX_i^{(k)}$ and $\mathcal{C}_{ik}$,
    then outputs a wide score $f_{wide}(\sparseConfigVectorNoEncloMath{},\sparseVarVectorNoEncloMath{}|\Theta_{s})$. 
    The \emph{wide} model uses a linear model over cross-product feature transformations to capture any occurrence of feature-pairs that commonly leads to effective visualizations. 
    
    \item \textbf{\emph{Deep} Vis. Rec. Model (Sec.~\ref{sec-deep-model}):} 
    The \emph{deep} model takes as input the dense features $\vd_x$ and $\vd_c$ of $\mX_i^{(k)}$ and $\mathcal{C}_{ik}$, then outputs a deep score $f_{deep}(\denseConfigVectorNoEncloMath{},\denseVarVectorNoEncloMath{}| \Theta_{d})$. 
    The \emph{deep} model uses dense features and non-linear transformations to generalize to unseen feature pairs that do not appear in the training set yet may lead to effective visualizations.
    
    \item \textbf{Training (Sec.~\ref{sec-learning-the-network}):} We describe the end-to-end \emph{training} of the wide-and-deep network model $\mathcal{M}$.
    The model and its parameters are learned using SGD over a sample of training visualizations from the corpus $\mathbcal{D} = \{\mX_i,\mathbb{V}_i\}_{i=1}^{N}$.
    
    \item \textbf{Scoring \& Recommending Visualizations via $\mathcal{M}$ (Sec.~\ref{sec-inference}):} 
    Given an entirely new unseen dataset $\mX_{\rm test}$, we then describe the inference procedure that uses $\mathcal{M}$ to score and recommend visualizations for the new dataset $\mX_{\rm test}$ of interest.
\end{itemize}

\noindent
Figure~\ref{fig-joint-modeling} illustrates how $\mathcal{M}$ operates at the granular level during both training (Sec.~\ref{sec-learning-the-network}) and inference (Sec.~\ref{sec-inference}): 
it predicts a numerical score $\hat{Y}_{ik}$ for each visualization $\mathcal{V}_{ik} = (\mX_i^{(k)}, \mathcal{C}_{ik})$ of a specific dataset $\mX_i$~\footnote{$\mX_i$ can be either a new unseen dataset $\mX_{\rm test}$, or a dataset from the training corpus $\mathbcal{D} = \{\mX_i,\mathbb{V}_i\}_{i=1}^{N}$}. 
The score $\hat{Y}_{ik}$ is given by 
\begin{equation}\label{eq:model-scoring}
    \hat{Y}_{ik} = \mathcal{M}(\mathcal{V}_{ik}) = f(\mX_i^{(k)}, \mathcal{C}_{ik}|\Theta) \in [0,1]
\end{equation}\noindent

\begin{table}
\caption{
Summary of notation.
Matrices are bold upright roman letters; vectors are bold lowercase letters.
}
\label{tab-summary-of-notation-approach}
\vspace{-2mm}
\centering 
\fontsize{8}{8.5}\selectfont
\setlength{\tabcolsep}{3pt} 
\def\arraystretch{1.3}
\begin{tabularx}{1.00\linewidth}{@{}p{30mm} X@{}} 
\toprule
$\mathbcal{D} = \{\mX_i,\mathbb{V}_i\}_{i=1}^{N}$ & a corpus of datasets $\{\mX_i\}_{i=1}^{N}$ and $N$ sets of visualizations $\{\mathbb{V}_i\}_{i=1}^{N}$ \\
$\mX_i = [\ldots\vx_{ij}\ldots]$ & an arbitrary dataset that has many attributes \\
$\vx$ & an attribute vector from an arbitrary dataset \\
$\mathbb{V}^{\star}_i = \mathbb{V}_i^{-} \cup \mathbb{V}_i $ & space of all possible visualizations that can be generated from $\mX_i$, also written as $\{
\ldots,\mathcal{V}_{ik},\ldots\}$ \\
$\mathbb{V}_i \subseteq \mathbb{V}^{\star}_i$ & set of positive visualizations (user-generated, observed) in $\mathbb{V}^{\star}_i$ \\
$\mathbb{V}_i^{-} \subseteq \mathbb{V}^{\star}_i$ & set of negative visualizations in $\mathbb{V}^{\star}_i$ \\
$\hat{\mathbb{V}}_i^{-} \subseteq \mathbb{V}_i^{-}$ & sampled negative visualizations from $\mathbb{V}_i^{-}$ \\
$\mathbcal{X}_i = \{\ldots\mX_i^{(k)},\ldots\}$ & space of attribute combinations for dataset $\mX_i$ \\
$\mathcal{V}_{ik} = (\mX_i^{(k)}\!,\mathcal{C}_{ik})$ 
& a visualization $\mathcal{V}_{ik}$ of dataset $\mX_i$ consisting of the attributes used in the visualization $\mX_i^{(k)}$ and a visualization configuration $\mathcal{C}_{ik}$ \\
$\mX_i^{(k)} \subseteq \mX_i$ & attribute combination in a visualization $\mathcal{V}_{ik}$ \\
$\mathbcal{C} = \{\ldots, \mathcal{C}_{ik}, \ldots\}$ & space of all visualization configurations \\
$\mathcal{C}_{ik} \in \mathbcal{C}$ & a visualization configuration \\
$\mathcal{M}$ & a visualization recommendation model \\
\midrule

$f(\mX_i^{(k)}, \mathcal{C}_{ik}|\Theta) = \hat{Y}_{ik}$ & scoring function of $\mathcal{M}$, parameterized by $\Theta$ on the input of $\mX_i^{(k)}$ and $\mathcal{C}_{ik}$ \\ 
$g(\cdot)$ & vector normalization  \\ 
$\phi_1$ & a concatenation operator \\
$\phi_k(\vs)$ & $k$-th cross-product transformation function \\
$\Theta$ & entire set of model parameters in $\mathcal{M}$ \\
$\mW_{s}^T$  & weight matrix of the wide model\\
$\vb$ & bias vector of the wide model\\
$\Theta_{s} =\{ \mW_{s}^T, \vb\}$ & model parameters for the wide model\\ 
$\{\textbf{W}_{d1},\ldots,\textbf{W}_{dL}^T\}$ & weight matrices of the deep model\\
$\{\textbf{b}_{d1}, \ldots, \textbf{b}_{dL}\}$ & bias vectors of the deep model\\
$\Theta_{d} =\{\textbf{W}_{d1},\!...,\textbf{b}_{d1},\!... \}$ &  model parameters for the deep model \\
$w_{wide}$ & weight to combine the wide score  $f_{wide}(\vs_c,\vs_{x}|\Theta_{s})$  \\
$w_{deep}$ & weight to combine the deep score $f_{deep}(\vd_c,\vd_{x}|\Theta_{d})$  \\
$\vs_x$  & sparse feature vector for an attribute combination $\mX_i^{(k)}$ \\
$\vs_c$  & sparse feature vector for a configuration $\mathcal{C}_{ik}$  \\
$\vs = \phi_1(\sparseConfigVectorNoEncloMath{}, \sparseVarVectorNoEncloMath{})$ & concatenated sparse features \\
$\vs^\prime$ & cross-product feature vector \\
$\vd_x$ & dense feature vector for attribute combination $\mX_i^{(k)}$ \\
$\vd_c$ & dense feature vector for a visualization configuration $\mathcal{C}_{ik}$  \\
$\vd = \phi_1(\denseConfigVectorNoEncloMath{}, \denseVarVectorNoEncloMath{})$ & concatenated dense features \\

\bottomrule
\end{tabularx}
\end{table} 

\subsection{Encoding the Input}
\label{sec-encode-input}
Every visualization can naturally be decomposed into the subset of attributes from the dataset $\mX_i$ and the visualization configuration, \ie, $\mathcal{V}_{ik} = (\mX_i^{(k)}, \mathcal{C}_{ik})$.
Since both $\mathcal{V}_{ik}$ and $\mX_i^{(k)}$ are specific to an arbitrary dataset $\mX_i$, the first step is to encode the input $\mX_i^{(k)}$ and $\mathcal{C}_{ik}$ into features in some shared space for the network. 

\subsubsection{Encode attributes into meta-features}
\label{sec-extract-meta-features}
Attributes from datasets in the training corpus $\{\mX_i\}_{i=1}^{N}$ are naturally from different domains and have fundamentally different characteristics such as their types, sizes, meanings, and so on.
This makes it fundamentally important to encode every attribute from any dataset in the corpus in a shared $K$-dimensional space where we can naturally characterize similarity between the attributes.
For this purpose, we leverage the meta-feature function $\Psi$ from Def.~\ref{def:meta-feature-function}.
Since $\Psi$ represents an attribute $\vx$ in a shared $K$-dimensional space, we apply $\Psi$ $\forall \vx \in \mX_i$.

We propose the meta-feature learning framework, as an instance of $\Psi$ for the network.
The framework has several components including nested meta-feature functions, attribute representation functions (where each of which can be used with the set of nested meta-feature functions), and so on. 
The framework is summarized in Table~\ref{table-meta-learning-framework}.
More formally, first we compute the meta-feature functions over different representations of the data as follows:
\begin{equation}\label{eq:meta-feature-different-data-representations}
\psi(\vx), \psi(p(\vx)), \psi(g(\vx)), \ldots,
\end{equation}\noindent
where $\psi(\vx)$ is the meta-features from $\vx$ directly, $\psi(p(\vx))$ are the meta-features from the probability distribution of $\vx$, and so on.
Next, given a partitioning (or clustering, binning) function $\Pi$ that divides a vector $\vx$ (or $p(\vx)$, $g(\vx)$) of values into $k$ partitions, we can derive meta-features for each partition as follows:
\begin{align}\label{eq:meta-features-of-partitions-single-partition-function}
\psi(\Pi_{1}(\vx)), \ldots, \psi(\Pi_{k}(\vx)), \\
\psi(\Pi_{1}(p(\vx))), \ldots, \psi(\Pi_{k}(p(\vx))), \\
\psi(\Pi_{1}(g(\vx))), \ldots, \psi(\Pi_{k}(g(\vx))) 
\end{align}
In the above, we use $\Pi_{k}$ to denote the $k$th partition of values from the partitioning function $\Pi$.
In this work, leverage multiple partitioning functions, and each can be used in a similar fashion as shown in Eq.~\ref{eq:meta-features-of-partitions-single-partition-function}.
All the meta-features derived from Eq.~\ref{eq:meta-feature-different-data-representations} and Eq.~\ref{eq:meta-features-of-partitions-single-partition-function} are then concatenated into a single vector of meta-features describing the characteristics of the attribute $\vx$.
More formally, the meta-feature function $\Psi : \vx \to \RR^{K}$ is defined as follows:
\begin{align}\label{eq:meta-features}
\!\!\!\!\!\Psi(\vx) = &\;\! \big[\psi(\vx), \psi(p(\vx)), \psi(g(\vx)),\!...,\!
                  \psi(\Pi_{1}(\vx)),\!...,\!\psi(\Pi_{k}(\vx)),\!..., \\
            &\; \psi(\Pi_{1}(p(\vx))),\!...,\!\psi(\Pi_{k}(p(\vx))),\!...,\! 
                  \psi(\Pi_{1}(g(\vx))),\!...,\!\psi(\Pi_{k}(g(\vx)))\big] \nonumber
\end{align}

As an example, given an attribute vector $\vx$ from any arbitrary dataset $\mX$, the first step is to derive many different data representations of $\vx$, e.g., using different normalization/scaling functions, probability distribution, log binning of $\vx$, etc. Then, we partition the values of each of the different representations of $\vx$ previously computed in Step 1 of Table~\ref{table-meta-learning-framework}. 
Now, for every different data representation of $\vx$ from Step 1 and every different partition of values from Step 2, we apply  meta-feature functions from Step 3 (see Table~\ref{table-meta-features}) over each one to get meta-features of $\vx$. 
Finally, we concatenate the meta-features from Step 3.
The resulting $\Psi(\vx)$ is a dense vector.
We denote it as $\vd_{x}$ - the \textit{meta-features}, a.k.a. the dense feature of the attribute $\vx$.
Without loss generality, we also normalize each meta-feature in $\vd_x$, by min-max scaling to scale each meta-feature value in $\vd_x$ to be between 0 and 1.
Our approach is agnostic to the precise meta-feature functions used, and is flexible to use with any alternative set of meta-feature functions.

\begin{table}
\centering
	\caption{
    Meta-feature framework for an attribute $\vx$.
	} 
	\label{table-meta-learning-framework}
	\vspace{-1mm}
	\small
	\begin{tabular}{l l  l@{} l @{} } 
		\toprule
		\textsc{Framework Components}  & \textsc{Examples} & 
		\\
		\midrule
		\textbf{1. Attribute representations} & $\vx$, $p(\vx)$, $g(\vx)$, $\ell_{b}\!(\vx)$, ...  &  &  \\ 
        \textbf{2. Partitioning values $\Pi$} & Clustering, binning, quartiles, ...  &  & \\
        \textbf{3. Meta-feature functions} $\psi$ & Statistical, info theoretic, ...  &  & \\

        \bottomrule
	\end{tabular}
\end{table}

\begin{table}[t]
\centering
	\caption{
	Summary of meta-feature functions for attribute. 
	The functions will be called from the learning framework in Table~\ref{table-meta-learning-framework}.
	Let $\vx$ denote an arbitrary attribute vector and $\pi(\vx)$ is the sorted vector of $\vx$.
	} 
	\label{table-meta-features}
	\vspace{-1mm}
	\small
	\begin{tabular}{@{}l l @{} H @{}H HHH  @{}} 
		\toprule
		\textbf{Name}  &
		\textbf{Equation} &
\textbf{Rationale} & \textbf{Variants} & 
\textbf{Representation} ($\vx$, ...) & 
\textbf{Attribute Type} (Q,C,T) &
\\
		\midrule
        Num. instances      &  $\abs{\vx}$ & Speed, Scalability         & $p/\abs{\vx}$, $log(\abs{\vx})$, $log(\abs{\vx}/p)$ \\ 
        Num. missing values &  $s$ & Imputation effects         &  \\
        Frac. of missing values &  $\nicefrac{\abs{\vx} - s}{\abs{\vx}}$ & Imputation effects  &  \\
        Num. nonzeros & $\mathtt{nnz}(\vx)$ & Imputation effects & \\
        Num. unique values & $\mathsf{card}(\vx)$ & Imputation effects & \\
        Density & $\nicefrac{\mathsf{nnz}(\vx)}{\abs{\vx}}$ & Imputation effects & \\
        
		\midrule
		$Q_1$, $Q_3$  & median of the $\abs{\vx}/2$ smallest (largest) values & $-$ & \\
        IQR  & $Q_3 - Q_1$ & $-$ & \\
        Outlier LB $\alpha \in \{1.5,3\}$ & $\sum_{i} \mathbb{I}(x_i < Q_1-\alpha IQR)$ & Data noisiness & \\
        Outlier UB $\alpha \in \{1.5,3\}$ & $\sum_{i} \mathbb{I}(x_i > Q_3+\alpha IQR)$ & Data noisiness & \\
        Total outliers $\alpha \in \{1.5,3\}$ & 
        $\sum_{i} \mathbb{I}(x_i \!<\! Q_1\!-\!\alpha IQR) + \sum_{i} \mathbb{I}(x_i \!>\! Q_3+\alpha IQR)$ 
        & Data noisiness & \\
        
        ($\alpha$std) outliers $\alpha \in \{2,3\}$        &  $\mu_{\vx} \pm \alpha \sigma_{\vx}$ & Data noisiness             & $o/\abs{\vx}$, lb, ub, total \\
        
        \midrule
        Spearman ($\rho$, p-val)  & $\mathsf{spearman}(\vx, \pi(\vx))$ & Sequential & \\
        Kendall ($\tau$, p-val) & $\mathsf{kendall}(\vx, \pi(\vx))$ & Sequential & \\
        Pearson ($r$, p-val) & $\mathsf{pearson}(\vx, \pi(\vx))$ & Sequential & \\

        \midrule

        Min, max & $\min(\vx)$, $\max(\vx)$ &  $-$ \\
        Range  & $\max(\vx) - \min(\vx)$ & Variable normality \\
        Median  & $\mathsf{med}(\vx)$ &  Variable normality \\
        
        Geometric Mean  & $\abs{\vx}^{-1} \prod_i x_i $ & Variable normality & \\
        Harmonic Mean & $\abs{\vx} / \sum_i \frac{1}{x_i}$ & Variable normality & \\
        Mean, Stdev, Variance  & $\mu_{\vx}$, $\sigma_{\vx}$, $\sigma^2_{\vx}$ & Variable normality & \\
		Skewness  & $\nicefrac{\mathbb{E}(\vx - \mu_{\vx})^3}{\sigma^3_{\vx}}$ & Variable normality & \\
        Kurtosis  & $\nicefrac{\mathbb{E}(\vx - \mu_{\vx})^4}{\sigma^4_{\vx}}$ & Variable normality &
        Fisher/Pearson, and bias/unbiased. \\
        HyperSkewness  & $\nicefrac{\mathbb{E}(\vx - \mu_{\vx})^5}{\sigma^5_{\vx}}$ & Variable normality & \\
        Moments [6-10]  & $-$ & Variable normality & \\
		k-statistic [3-4]  & $-$ & Variable normality &  \\
        \midrule
        Quartile Dispersion Coeff. & $\frac{Q_3-Q_1}{Q_3+Q_1}$ & Dispersion & \\
        Median Absolute Deviation & $\mathsf{med}(\abs{\vx - \mathsf{med}(\vx)})$ & Dispersion & \\
        Avg. Absolute Deviation & $\frac{1}{\abs{\vx}} \ve^T\!\abs{\vx - \mu_{\vx}} $ & Dispersion & \\
        
        Coeff. of Variation & $\nicefrac{\sigma_{\vx}}{\mu_{\vx}}$ & Dispersion & \\
        
        Efficiency ratio & $\nicefrac{\sigma^2_{\vx}}{\mu^2_{\vx}}$ & Dispersion & \\
        Variance-to-mean ratio & $\nicefrac{\sigma^2_{\vx}}{\mu_{\vx}}$ & Dispersion & \\
        \midrule
        
        Signal-to-noise ratio (SNR)  & $\nicefrac{\mu^2_{\vx}}{\sigma^2_{\vx}}$ & Noisiness of data  & \\
        Entropy & $H(\vx) = -\sum_{i} \; x_i \log x_i$ & Variable Informativeness  & \\
        Norm. entropy & $\nicefrac{H(\vx)}{\log_2 \abs{\vx}}$ & Variable Informativeness  & \\
        Gini coefficient & $-$ & Variable Informativeness & \\
        \midrule
        Quartile max gap & $\max (Q_{i+1} - Q_{i}) $ & Dispersion & \\
        Centroid max gap & $\max_{ij} |c_{i} - c_{j}| $ & Dispersion & \\
        
        \midrule
        Histogram prob. dist. & $\vp_h = \frac{\vh}{\vh^T\ve}$ (with fixed \# of bins) & - & \\

        \bottomrule
	\end{tabular}
\end{table}

We obtain $\Psi(\vx)$ as the meta-features for each attribute $\vx$. 
An attribute combination $\mX_i^{(k)}$ usually has more than one attributes, whose meta-features need to be combined to get an overall dense feature $\vd_{x}$. 
We concatenate all of meta-features $\vd_{x_{ij}}$ from each attribute $\vx_{ij} \in \mX_i^{(k)}$ to get the overall dense feature $\vd_{x}$, written as 
\begin{equation}
\begin{aligned}
&\vd_{x} = \phi_1(\ldots \vd_{x_{ij}} \ldots) =  
\begin{bmatrix} \vdots \\ \vd_{x_{ij}} \\ \vdots \end{bmatrix} 
\end{aligned}
\end{equation}\noindent

See Figure~\ref{fig-joint-modeling} for an example. 
It has two attributes selected, i.e. $\mX_i^{(k)} = [\,\vx_r \; \vx_s\,]$, which results in two dense vectors $\vd_{x_{r}}$ and $\vd_{x_{s}}$.
The overall dense feature $\vd_{x}$ therefore is $\vd_{x} = \phi_1(\vd_{x_{r}}, \vd_{x_{s}})$.

\subsubsection{Visualization Configuration Embedding}
The space of all possible configurations $\mathbcal{C}= \{\ldots, \mathcal{C}_{ik}, \ldots\}$ is a shared set for all visualizations from any dataset. 
Let $\mathcal{C}_{ik}$ denote one configuration in the space.
Like all other configurations, although we denote $\mathcal{C}_{ik}$ as the configuration of the visualization of our interest, where $\mathcal{V}_{ik} = (\mX_i^{(k)}, \mathcal{C}_{ik})$, this configuration is independent from any dataset or visualization (Property~\ref{prop:visualization-configuration-independent}).
It is possible to learn embeddings for all configurations in $\mathbcal{C}$ and use the embeddings to encode $\mathcal{C}_{ik}$.

\begin{Definition}[Configuration Embedding Function] \label{def:configuration-embedding-function}
Let $E$ denote a configuration embedding function that maps a configuration $\mathcal{C}_{ik}$ to a shared $K$-dimensional embedding space such that the embedding $\mathcal{H}(\mathcal{C})$ captures the important characteristics of $\mathcal{C}_{ik}$ and can be learned along with the model $\mathcal{M}$.
More formally, 
\begin{equation}
\label{eq:config-embedding-single-config}
    \mathcal{H} : \mathcal{C}_{ik} \to \RR^{K}
\end{equation}\noindent
Further, given the space of all visualization configurations $\mathbcal{C}$ of size $M = |\mathbcal{C}|$, then we obtain a $K$-dimensional embedding matrix for all visualization configurations as $\mathcal{H} (\mathbcal{C})$
\begin{equation}\label{eq:config-embedding-all-config}
    \mathcal{H}  : \mathbcal{C} \to \RR^{K \times M}
\end{equation}\noindent
\end{Definition}
We denote $\mathcal{H} (\mathcal{C}_{ik})$ as the dense feature $\vd_c$ of the configuration $\mathcal{C}_{ik}$, i.e. $\vd_c = \mathcal{H}(\mathcal{C}_{ik})$.
$\mathcal{H}$ works as follows:
Suppose we are scoring visualizations for an arbitrary dataset, one visualization is $\mathcal{V}_{ik} = (\mX_i^{(k)}, \mathcal{C}_{ik})$. 
We first abstract the configuration $\mathcal{C}_{ik}$ from $\mathcal{V}_{ik}$, and look up the positional identity of $\mathcal{C}_{ik}$ in $\mathbcal{C}$.
Then, we one-hot encode the identity $\mathcal{C}_{ik}$ and apply configuration embedding function $\mathcal{H}$ to the one-hot encoding. 
This gives a $k$-dimensional dense feature $\vd_c$, written as 
\begin{equation}
\vd_c = \mathcal{H}(one\_hot(\mathcal{C}_{ik}))
\end{equation}\noindent
Note that $\mathcal{H}$ is learnable with the model $\mathcal{M}$.

\subsubsection{Complement Dense Features with Sparse Features}
Up so far, both the configuration embedding vector $\vd_c$ and attribute meta-features $\vd_x$ are dense features (vectors in real-value). 
On the other hand, our approach wants to capture some frequent feature patterns about the attribute combination $\mX_i^{(k)}$ and the configuration $\mathcal{C}_{ik}$ that commonly lead to effective visualizations. 
The frequent feature patterns can be best expressed through sparse features, i.e. \textit{whether this visualization has the feature(s) X or not}. 
For example, scatterplot is generally more effective to visualize attributes that have many rows, than line charts and bar charts. 
If a visualization $\mathcal{V}_i^{(k)} = (\mX_i^{(k)}, \mathcal{C}_{ik})$ has sparse features indicating that the number of rows in one attribute of $\mX_i^{(k)}$ is larger than 50 and the configuration $\mathcal{C}_{ik}$ is about scatterplot, our model $\mathcal{M}$ should be able to assign a high score to this visualization and consider it as effective. 
Therefore, we create the set of sparse features $\vs_x$ and $\vs_c$ to complement the dense features $\vd_x$ and $\vd_c$, which will also be used as the input to $\mathcal{M}$.

There are many choices to create sparse features $\vs_x$ and $\vs_c$.
For example, one simple option to get the sparse feature $\vs_x$ for attribute combination $\mX_i^{(k)}$ is to bin-bucket the dense features $\vd_x$. 
Recall that the dense feature (i.e. meta-features) $\vd_x$ is a vector normalized in each dimension. 
We could bin-bucket each dimension of the normalized meta-features $\vd_x$ into a fixed number of $n$-bins within the range of $[0,1]$.
Each bin has an equal width of $\frac{1}{n}$.
Another option to get the sparse features from $\vd_x$ is to first cluster each dimension from $\vd_x$ of all seen visualizations, then one-hot encode the cluster identity for the value in each dimension of the dense feature $\vd_x$.
Our wide-and-deep network is agnostic to the actual option and the precise meta-features that are being used in $\vd_x$. 

To get the sparse feature $\vs_c$ from a configuration embedding vector $\vd_c$, one option is to use the original one-hot sparse vector as its sparse feature.
Another option is to one-hot encode each pair of field and value that appears in the configuration $\mathcal{C}_{ik}$. 
For example, we could assign a value of 1 to one dimension of $\vs_{c}$ for a configuration $\mathcal{C}_{ik}$, if the configuration $\mathcal{C}_{ik}$ satisfies a specific pair of field and value, such as ``\texttt{marker.symbol = circle}.'' 

\subsection{The Wide Model}
\label{sec-wide-model}
The \emph{Wide} model is a linear model over the set of sparse features \sparseConfigVector{} and \sparseVarVector{}. 
The goal of leveraging sparse features is to capture any occurrence of feature-pairs that commonly lead to effective visualizations in the training corpus. 
As an example, if the corpus $\mathbcal{D} = \{\mX_i,\mathbb{V}_i\}_{i=1}^{N}$ has many visualizations that use scatterplot with default point size and point color to visualize two quantitative attributes with more than 50 rows,~\footnote{Note on this pattern, which will be reused in Sec.~\ref{sec-deep-model} for motivations of the \emph{Deep} model.} a fully-trained model $\mathcal{M}$ should be able to pick up this pattern: when a new dataset comes in, which has over 50 rows and at least two quantitative attributes in similar characteristics, $\mathcal{M}$ would be able to generate, score, and recommend a similar-style scatterplot that visualizes over a subset of two quantitative attributes.

First, we concatenate \sparseConfigVector{} and \sparseVarVector{} into one single sparse vector $\vs$ where $\phi_1$ is a concatenation operator.
 
\begin{equation}
\begin{aligned}
	\vs &= \phi_1(\sparseConfigVectorNoEncloMath{}, \sparseVarVectorNoEncloMath{}) = 
	\begin{bmatrix} \sparseConfigVectorNoEncloMath{} \\ \sparseVarVectorNoEncloMath{} \end{bmatrix} 
\end{aligned}
\end{equation}

Next, we augment $\vs$ with \textit{cross-product features} from $\vs$, denoted as $\vs^\prime$. 
\textit{Cross-product features} $\vs^\prime$ captures co-occurrences of some specific features in the original $\vs$. 
Formally, it is calculated as the concatenation of values from a set of cross-product transformation functions.
\begin{equation}
\begin{aligned}
\vs^\prime = \{\ldots, \phi_k(\vs), \ldots\} \\
\end{aligned}
\end{equation}
where $\phi_k(\vs)$ is the $k$-th cross-product transformation function. 
The operator $\phi_k(\cdot)$ checks whether a few selected dimensions in $\vs$ are all 1, written as
\begin{equation}
    \phi_k(\vs) = \prod_{i=1}^{|\vs|} \vs_i^{t_{ki}}, \quad t_{ki} \in \{0,1\}
\end{equation}
where $t_{ki}$ is a boolean value indicating whether or not the $k$-th cross-product transformation function $\phi_k(\vs)$ ``cares'' about the $i$-th feature of $\vs$. 
For example, suppose $\phi_k(\cdot)$ checks whether a visualization satisfies  
(1) the entropy of its first attribute is in the range of $[0.2, 0.4)$ and (2) its configuration is configuration no.3.
The cross-product feature $\phi_k(\vs)$ is 1 if and only $\vs$ has feature dimensions of {\small\texttt{entropy-1st-var-bucket=2}} and {\small\texttt{config-bucket=3}} both as 1.

Finally, the sparse feature $\vs$ and the cross-product transformed feature $\vs^\prime$ get concatenated using the concatenation operator $\phi_1$, then go through a linear transformation, to get the wide score.
More formally, the wide score is 
\label{sec:wide-model}
\begin{equation} \label{eq:func-wide}
f_{wide}(\sparseConfigVectorNoEncloMath{},\sparseVarVectorNoEncloMath{}|\Theta_{s}) =  \mW_{s}^T[\vs, \vs^\prime] + \vb_s  
\end{equation}
where $\mW_{s}^T$ and $\vb$ denote the weight matrix and the bias vector for the wide model.
$\Theta_{s} =\{ \mW_{s}^T, \vb\}$ denotes the entire set of parameters in the wide model. 
The wide score is a numeric value, i.e. satisfies $f_{wide}(\sparseConfigVectorNoEncloMath{},\sparseVarVectorNoEncloMath{}|\Theta_{s}) \in \RR$.

\subsection{The Deep Model}
\label{sec-deep-model}
The \emph{Deep} model uses dense features and non-linear transformations to generalize to feature pairs that do not frequently appear in the training set yet may lead to effective visualizations with a good rationale.
Suppose the same corpus $\mathbcal{D} = \{\mX_i,\mathbb{V}_i\}_{i=1}^{N}$ as in Sec.~\ref{sec-wide-model} not only has the frequently-observed pattern about scatterplots, but also a few scatterplots with half point size that visualize two quantitative attributes with hundreds of rows, a fully-trained model $\mathcal{M}$ should be able to generalize from this. 
When a new dataset with thousands of rows and at least two quantitative attributes comes in, $\mathcal{M}$ would be able to generate, score, and recommend a scatterplot that preferably has smaller point size to visualize a subset of two quantitative attributes.

The deep model works by first concatenating the two dense features \denseConfigVector{} and \denseVarVector{} into an intermediate vector $\vd$, such that it incorporates the information from both the configuration and the attribute combination. 
\begin{equation}
\begin{aligned}
	\vd &= \phi_1(\denseConfigVectorNoEncloMath{}, \denseVarVectorNoEncloMath{}) = 
	\begin{bmatrix} \denseConfigVectorNoEncloMath{} \\ \denseVarVectorNoEncloMath{} \end{bmatrix} \\
\end{aligned}
\end{equation}

The concatenated vector $\vd$ are then fed into a total of $L$ hidden layers (standard MLP layers).
The initial layer starts with $\vd$.
At the $k$-th layer, an intermediate vector $\vd_{k-1}$ from the previous layer $(k-1)$ go through non-linear transformations with the model parameter $\textbf{W}_{dk}$ and the activation function $a_{k-1}$.
The activation function $a_{k-1}$ could either be the rectified linear unit (ReLU) or the sigmoid function. 
This design offers greater flexibility to model feature interaction. 
The last layer gives the output from the deep model, as the deep score $f_{deep}(\denseConfigVectorNoEncloMath{},\denseVarVectorNoEncloMath{}| \Theta_{d})$. 
Formally, it can be written as 
\begin{equation}
\begin{aligned}
\vd_0 &= \vd \\
\vd_1 &= a_1(\textbf{W}_{d1}^T \vd_0 + \textbf{b}_{d1}), \\
	&...... \\
	\vd_{L-1} &= a_{L-1}(\textbf{W}_{d(L-1)}^T \vd_{L-2} + \textbf{b}_{d(L-1)}), \\
f_{deep}(\denseConfigVectorNoEncloMath{},\denseVarVectorNoEncloMath{}| \Theta_{d}) &= a_{L}(\textbf{W}_{dL}^T \vd_{L-1} + \textbf{b}_{dL}),
\end{aligned}
\end{equation}
where $\{\textbf{W}_{d1},\ldots,\textbf{W}_{dL}^T\}$ and $\{\textbf{b}_{d1}, \ldots, \textbf{b}_{dL}\}$ denote the weight matrices and the bias vectors for the deep model, and $\Theta_{d} =$
$\{\textbf{W}_{d1},...,\textbf{W}_{dL}^T,$ 
$\textbf{b}_{d1},...,\textbf{b}_{dL}\}$ denotes the entire set of parameters in the deep model. 
The deep score is a numeric value, i.e. satisfies $f_{deep}(\denseConfigVectorNoEncloMath{},\denseVarVectorNoEncloMath{}| \Theta_{d}) \in \RR$.

\subsection{Training the Network}
\label{sec-learning-the-network}
Previous two sections describe the set of model parameters $\Theta$ that constitutes the wide-and-deep network and that goes into Eq.~\ref{eq:model-scoring} $\hat{Y}_{ik} = \mathcal{M}(\mathcal{V}_{ik}) = f(\mX_i^{(k)}, \mathcal{C}_{ik}|\Theta)$.
In this section, we elaborate upon Def.~\ref{def:ML-based-vis-rec-model-learning-training} to show how to optimize the wide-and-deep network parameters $\Theta$ with a probabilistic approach~\cite{he2017neural}. 

The training corpus $\mathbcal{D} = \{\mX_i,\mathbb{V}_i\}_{i=1}^{N}$ has a set of datasets $\{\mX_i\}_{i=1}^{N}$.
Each dataset $\mX_i$ has a set of positive visualizations $\mathbb{V}_i$, which we also complement a sampled set of negative visualizations $\hat{\mathbb{V}}_i^{-}$ (as in Def.~\ref{def:negative-vis-sampling}). 
The set of training visualizations for $\mX_i$ is then $\mathbb{V}_i \cup \hat{\mathbb{V}}_i^{-}$.
In other words, during training, each visualization $\mathcal{V}_{ik}$ comes from an arbitrary dataset $\mX_i$ 
and has a binary ground-truth label $Y_{ik} \in \{0, 1\}$.
A label of $1$ indicates a positive visualization, i.e. $\mathcal{V}_{ik} \in \hat{\mathbb{V}}_i$.
Hence, the visualization is generated by the user.
Label $0$ indicates a negative (non-relevant) visualization, i.e. $\mathcal{V}_{ik} \in \hat{\mathbb{V}}_i^{-}$.
Non-relevant visualizations are sampled from the space of all visualizations that belong to the dataset $\mX_i$. 
Def.~\ref{def:negative-vis-sampling} and Section~\ref{sec-framework-training-negative-sampling} discuss more details about how we sample non-relevant visualizations to support training.

Our goal is to have the model score $\hat{Y}_{ik} \in [0,1]$ of each training visualization $\mathcal{V}_{ik}$ as close as possible to its ground-truth label $Y_{ik}$. 
We train the model by optimizing the likelihood of model scores rounding up to match the ground-truth labels, for all visualizations throughout the entire corpus $\mathbcal{D}$  consisting of $N$ datasets $\{\mX_i\}_{i=1}^N$.
Eq.~\ref{eq:likelihood} shows the calculation of the likelihood: for each dataset $\mX_i$ we have the set $\mathbb{V}_i \cup  \hat{\mathbb{V}}_i^{-} = \{\ldots, (\mX_i^{(k)},\!\mathcal{C}_{ik}),\ldots\}$ of training visualizations where each visualization $(\mX_i^{(k)}, \mathcal{C}_{ik}) \in \mathbb{V}_i \cup  \hat{\mathbb{V}}_i^{-}$ consists of the configuration $\mathcal{C}_{ik} \in \mathbcal{C}$ and the subset of attributes $\mX_i^{(k)}$ from the dataset $\mX_i$.
\begin{equation}\label{eq:likelihood}
\!\!
p(\hat{\mathbb{V}}_i^{-}, \mathbb{V}_i| \Theta) = \!
\prod_{(\mX_i^{(k)}\!, \mathcal{C}_{ik})\in\mathbb{V}_i} \!\hat{Y}_{ik} 
\prod_{(\mX_i^{(k)}\!,\mathcal{C}_{ik})\in \hat{\mathbb{V}}_i^{-}} \!\!\Big(1 - \hat{Y}_{ik}\Big), 
\;\;
\text{for } i=1,\ldots,N
\end{equation}
 
The closer that $\hat{Y}_{ik}$ is to the ground-truth label $Y_{ik}$, the better. 
Taking the negative log of the likelihood in Eq.~\ref{eq:likelihood} and summing over all datasets $\{\mX_i\}_{i=1}^N$ give us the loss $L$. 
\begin{equation}
\label{eq:objective}
\begin{aligned}
L &= \sum_{i=1}^{N}\Big(-\sum_{(\mX_i^{(k)}, \mathcal{C}_{ik})\in\mathbb{V}_i}\log \hat{Y}_{ik} - \sum_{(\mX_i^{(k)}, \mathcal{C}_{ik})\in \hat{\mathbb{V}}_i^{-}}\log (1 - \hat{Y}_{ik})\Big) \\
&= - \sum_{i=1}^{N}\sum_{(\mX_i^{(k)}, \mathcal{C}_{ik})\in\mathbb{V}_i \cup \hat{\mathbb{V}}_i^{-}} Y_{ik}\log \hat{Y}_{ik} + (1 - Y_{ik}) \log (1 - \hat{Y}_{ik})
\end{aligned}
\end{equation}
We minimize the objective function through stochastic gradient descent (SGD) to update the model parameters $\Theta$ in $\mathcal{M}$.

\subsection{Inference}
\label{sec-inference}
Given the \emph{trained} wide-and-deep visualization recommendation model $\mathcal{M}$ from Section~\ref{sec-learning-the-network}, we now describe the inference procedure for scoring and recommending visualizations from an arbitrary new dataset of interest.
Recall from Section~\ref{sec-problem-formulation} that the set of visualizations to recommend depends entirely on the dataset of interest, that is, the set of visualizations for one dataset is guaranteed to be completely disjoint for another dataset.
As illustrated in the lower part of Figure~\ref{fig-joint-modeling}, given an arbitrary dataset $\mX_{\rm test}$ selected or uploaded by an arbitrary user, we generate the space of visualizations $\mathbb{V}^{\star}_{\rm test} = \{\ldots, \mathcal{V}_{\rm test}^{(k)}, \ldots\}$ through Def.~\ref{def:space-of-visualizations} process, where each visualization $\mathcal{V}_{\rm test}^{(k)}$ consists of a subset of attributes $\mX_{\rm test}^{(k)}$ from the dataset $\mX_{\rm test}$ and a configuration $\mathcal{C} \in \mathbcal{C}$, i.e. $\mathcal{V}_{\rm test}^{(k)}=(\mX_{\rm test}^{(k)}, \mathcal{C})$. 

Each visualization $\mathcal{V}_{\rm test}^{(k)}$ will be fed into $\mathcal{M}$ for scoring. 
First, we encode the configuration $\mathcal{C}$ into the sparse feature $\vs_c$ and the dense feature (configuration embedding) $\vd_c$. 
Given the attribute combination $\mX_{\rm test}^{(k)}$, we derive the meta-feature $\vd_{x_{i}}$ for each attribute $\vx_i \in \mX_{\rm test}^{(k)}$. 
The meta-features get concatenated to get an overall dense feature on attribute combination, as $\vd_{x}$.
Bin-bucking $\vd_{x}$ gives the sparse feature on attribute selection, as $\vs_x$. 
The features go through the network where the wide model in Sec.~\ref{sec-wide-model} and the deep model in Sec.~\ref{sec-deep-model} transform them into a wide score and a deep score, denoted as $f_{wide}(\vs_c,\vs_{x}|\Theta_{s})$ and $f_{deep}(\vd_c,\vd_{x}| \Theta_{d})$ respectively, where $\Theta_{s}$ and $\Theta_{d}$ are model parameters in the wide model and the deep model.
The network then weighs the two score vectors with respective weights, denoted as $w_{wide}$ and $w_{deep}$, to get a final score $\hat{Y}_{test,k} \in [0,1]$ as follows,
\begin{align}
\hat{Y}_{test,k}&= f(\mX_{\rm test}^{(k)}, \mathcal{C}|\Theta) \\
&= \sigma(w_{wide}  f_{wide}(\vs_c,\vs_{x}|\Theta_{s}) +  w_{deep} f_{deep}(\vd_c,\vd_{x}| \Theta_{d}))  \nonumber
\end{align}\noindent
where $f_{wide}(\vs_c,\vs_{x}|\Theta_{s})$ is the wide score and $f_{deep}(\vd_c,\vd_{x}| \Theta_{d})$ is the deep score.
$w_{wide}$ and $w_{deep}$ are two real values, i.e. $w_{wide}\in \RR$ and $w_{deep}\in \RR$.
The entire set of parameters $\Theta$, including $\Theta_{s}$, $\Theta_{d}$, $w_{wide}$ and $w_{deep}$ are learned through backward propagation as described in Section~\ref{sec-learning-the-network}.

We repeat the above to score all possible visualizations in $\mathbb{V}^{\star}_{\rm test}$, and then recommend top visualizations based on the prediction scores. 
This process is consistent with Def.~\ref{def:ML-based-vis-rec}.
We expand more details about the evaluation of this process in Section~\ref{sec-test-framework}. 

\section{Evaluation Framework}
\label{sec:eval-framework-MLVisRec}
One of the contributions in this work is the proposed evaluation framework for end-to-end \emph{ML-based} visualization recommendation systems.
This framework serves as a fundamental basis for systematically evaluating ML-based visualization recommender systems, including our own model and those that arise in the future.
We first summarize the differences that make it infeasible to use traditional techniques for evaluation of ML-based visualization recommendation systems, which motivates the need for such a framework, and then discuss each component of the evaluation framework.

\subsection{Motivation}
\label{sec-framework-motivation}
Since the visualization recommendation problem is fundamentally different from the traditional recommendation problem (\ie, recommending items to users), we are unable to leverage the same commonly used evaluation techniques.
We summarize some of these fundamental differences below, which motivate the need for the proposed evaluation framework.
\begin{itemize}[leftmargin=*]
    \item \textbf{Visualization Complexity (Sec.~\ref{sec-framework-corpus})}: While traditional recommender systems have a simple object to recommend such as an item, ML-based visualization recommendation models must learn from a far more complex visualization object consisting of a subset of attributes from an arbitrary dataset, and a set of design choices.
    
    \item \textbf{No Shared Recommendation Space (Sec.~\ref{sec-framework-corpus})}: 
    Visualizations recommended for one dataset cannot be recommended for another dataset. 
    Hence, there is no shared space of visualizations 
    for learning better recommender models.
    
    \item \textbf{Generate On-The-Fly (Sec.~\ref{sec-framework-training-negative-sampling}-\ref{sec-test-framework})}: 
    Set of visualizations to recommend are generated on-the-fly for a specific unseen dataset of interest, as opposed to already existing and being common to all users as is the case for traditional recommender systems.
    For instance, when a user uploads a new dataset, ML-based visualization recommender systems must generate relevant visualizations that are only applicable for the user-specific dataset of interest.

    \item \textbf{Dynamic \& Dataset Dependent Vis. Space (Sec.~\ref{sec-framework-training-negative-sampling}-\ref{sec-test-framework})}: 
    Space of visualizations to score and recommend is dynamic, completely \emph{dependent} on the individual dataset of interest, and exponential in the number of attributes and possible design choices.
\end{itemize}

\subsection{Corpus: Datasets and Visualizations}
\label{sec-framework-corpus}
In most traditional recommender systems that recommend items to users, there is a \emph{single shared set of items} (\eg, movies on Netflix, products on Amazon, etc).
However, in visualization recommendation, there is not a shared set of visualizations to recommend to users, 
as it depends entirely on the dataset of interest.
Hence, if we have $N$ datasets, then there are $N$ completely disjoint sets of visualizations that can be recommended.
However, in visualization recommendation, we begin with a general corpus consisting of datasets and relevant visualizations.
Each dataset has a set of relevant visualizations that are exclusive to the dataset. 
Moreover, each visualization only uses a small subset of attributes from the dataset.
There can be attributes in the dataset that are never used in a visualization.
While the goal of traditional recommender systems is typically to recommend items (from a specific dataset) to users, in visualization recommendation, the goal is to learn a model to score and ultimately recommend visualizations that are generated for a specific unseen dataset.
Therefore, the model learned in visualization recommender systems must be able to generalize for use in scoring visualizations generated from any unseen dataset in the future.

Every new dataset gives rise to an exponential amount of possible visualizations. 
This makes this recommendation problem extremely challenging.
In addition to the exponential space of visualizations that one must search for just a single dataset, the visualization search space is also completely disjoint from the search space of another arbitrary dataset as shown in Lemma~\ref{lem:vis-space-data-dependence}.
Therefore, given an available corpus that consist of datasets and relevant visualizations, our framework first splits the corpus by datasets into various sets required for training, validation, and testing.
For datasets in the testing set, Section~\ref{sec-eval-metric} discusses how to apply evaluation metrics to a number of ranked lists, where each list has recommended visualizations that are tied to one test dataset. 

To learn a ML-based visualization recommendation model within our framework, we can use any available corpus as long as it has a set of datasets and relevant user-created visualizations that use a subset of attributes from the datasets.
Notably, the corpus can be visualizations and datasets from a variety of different sources, \eg, they can be visualizations and datasets collected from the web or even a visual analytics platform such as Tableau.

\subsection{Training from the Corpus}
\label{sec-framework-training-negative-sampling}
The next step is to create a training set from the corpus of datasets and visualizations.
For example, the wide-and-deep network approach addresses the issue of the dynamic space of visualizations by creating visualizations that come from a combination of attributes and a visualization configuration.
However, the framework would generalize to other approaches of visualization recommendation that may have a different way extracting a visualization instance.
Given a single dataset 
from the corpus that has a set of relevant visualizations, we construct positive (relevant) visualizations as in Figure~\ref{fig-plotly-schema-example}, and complement with ``negative'' (or non-relevant) visualizations.

While it is intuitive to think that non-relevant visualizations in our problem are visualizations that users do not create, such set of non-relevant instances does not naturally exist in the corpus.
The corpus only contains visualizations that users do create.
Our framework needs to compute non-relevant visualizations on-the-fly from the dynamic space of visualizations that depends on each dataset.
In other words, given a different dataset, there is a different set of non-relevant visualizations since the underlying data in the actual visualizations is different.
Our framework follows Def.~\ref{def:negative-vis-of-a-dataset} to achieve that. 
Moreover, the space of non-relevant visualizations is typically exponential 
with a size that easily exceeds several thousands.
It is difficult to train with a large number of non-relevant visualizations along with a much smaller set of relevant visualizations. 
Our framework follows Def.~\ref{def:negative-vis-sampling} to uniformly sample a fixed number 
of non-relevant visualizations from the same dataset.
Our framework also welcomes other ways of sampling non-relevant visualizations, e.g. drawing non-uniform samples from the pool of non-relevant visualizations (e.g. based on the popularity of the configuration in a visualization, or biased to sample  most-similar or least-similar non-relevant visualizations to relevant visualizations).

\subsection{Testing and Deployment}
\label{sec-test-framework}
Now, we discuss how a visualization recommendation model $\mathcal{M}$ learned from the training set of relevant and non-relevant visualizations can be used for testing and deployment.
Given a new or selected held-out dataset from the corpus, the model outputs a list of recommended visualizations for the specific dataset.
Different from traditional recommender systems where the space of items are shared for all users (including new users), our framework generates a ranking of visualizations to recommend with on-the-fly, which are dependent to the dataset.
As the entire space of visualizations for each dataset can be large, negative sampling of non-relevant visualizations allow us to test on more datasets efficiently and derive evaluation metrics without losing statistical rigor. 
We describe the evaluation metrics for our visualization recommendation problem in Sec.~\ref{sec-eval-metric}.
When the model gets deployed and tested on a new dataset, the recommended visualizations are selected from the entire space of visualizations for that dataset (as in Def.~\ref{def:space-of-visualizations}).

\subsection{Evaluation Metrics}
\label{sec-eval-metric}
Evaluating the quality of the ranking of visualizations (recommendations) given by the learned model has its unique challenges.
We summarize the challenges and propose a suitable evaluation metric.

In traditional recommender systems, there is a shared global set of items to recommend to any user.
However, in visualization recommendation, the set of visualizations to be recommended is generated on-the-fly based on the dataset of interest.
Since each dataset gives rise to a new set of visualizations that can be recommended, we therefore must evaluate the quality of ranking for each individual dataset and explicitly account for the different space of visualizations being ranked for every different dataset $\mX_i$.
Therefore, the standard ranking metrics such as nDCG cannot be used directly for visualization recommendation.
This evaluation must be performed completely independent of any other dataset in the corpus.
For instance, in traditional recommender systems, we simply use a model to infer scores for every item since all items are shared by all users.
However, in visualization recommender systems, suppose we want to evaluate whether the model can rank actual relevant/positive visualizations from a held-out test dataset highly, then we have to generate all possible visualizations for the specific dataset, and then compute the ranking metric over this set of visualizations independently of other visualizations from other datasets.
As such, we have to repeat this process for every dataset, correct for the difference in space, and then average the result.

Furthermore, the number of possible visualizations the ranking is computed over depends entirely on the dataset and the number of attributes in it.
Therefore, the difficulty of the visualization ranking problem varies based on the dataset, and more specifically, the number of attributes in that dataset.
For instance, it is easy to score a high nDCG for a dataset with only two attributes as opposed to one with hundreds.

For these reasons,the  traditional ranking metrics (\eg, nDCG) are not appropriate for the visualization recommendation problem and give incorrect and misleading results.
This leads us to propose a modified version of nDCG that can be used for evaluation of ML-based visualization recommendation models.
As an aside, other evaluation metrics can also be corrected in a similar fashion.
Given $N$ test datasets along with $N$ sets of held-out positive visualizations $\{\mathbb{V}_{i}\}_{i=1}^N$, then we propose a modified nDCG defined formally as:
\begin{align}\label{eq-our-ndcg}
    nDCG@K =& \frac{1}{N} \sum_{i=1}^{N} 
    \frac{1}{Z^{K}_{i}}
    \sum_{j=1}^{K} \frac{2^{Y_{ij}}-1}{log_2(j+1)} \\
    Z^K_i =& \sum_{j=1}^{\min(K, |\mathbb{V}_i|)} \frac{1}{log_2(j+1)}  \label{eq-our-ndcg-normalizer}
\end{align}
where $j$ is the rank, $Y_{ij} \in \{0,1\}$ is the ground-truth label (relevant/irrelevant) of the visualization at position $j$ in the ranking of visualizations for dataset $\mX_i$, 
and $Z_{i}^{K}$ is the normalization factor for dataset $\mX_i$.
The dataset-dependent normalization factor $Z^{K}_{i}$ ensures that a perfect ranking for our visualization recommendation problem receives a perfect score of 1.
This is required since each dataset $\mX_i$ may have a different number of positive visualizations $|\mathbb{V}_i|$, that is, for any arbitrary two datasets $\mX_i$ and $\mX_j$, $|\mathbb{V}_i| \not= |\mathbb{V}_j|$. 
The perfect ranking recommends all (but no more than $K$) positive visualizations at the top. 
Our modified \textit{nDCG} emphasizes the quality of the visualization ranking at the top of the list of recommended visualizations for each dataset $\mX_i$ since $1 / \log2(j + 2)$ decreases quickly and then asymptotes to a constant as $j$ increases.
Therefore, a good end-to-end learning-based visualization recommender system must be able to give up some of its performance at the bottom of the list of recommended visualizations to improve the performance at the top.
In Sec.~\ref{sec:exp-quant-results}, we use the proposed \textit{nDCG} metric from Eq.~\ref{eq-our-ndcg} up to the 20th position.

\begin{table*}[t!]
\begin{center}
\caption{Training Corpus Statistics.
\textsc{\# Config/Dataset} denotes the average number of configurations used by each dataset.
\eat{Add \# of meta-features?}
}
\label{tab:dataset-stats} 
\begin{tabular}{cccccc}
\toprule
\textsc{\#Datasets} & \textsc{\#Vis. Configs} & \textsc{\#Attributes} & 
\textsc{\#Visualizations} &
\textsc{\#Attribute/Dataset}& \textsc{\# Vis. Configs/Dataset} \\ 
\midrule
925 & 60 & 11,778 & 4,865 & 11.93 & 5.89 \\ \bottomrule
\end{tabular}
\end{center}
\end{table*}

\begin{table*}[t!]
\centering
    \caption{Quantitative Results for Visualization Recommendation. 
    See text for discussion.
    }
    \label{tab:rank-one-seen-among-unseen}
    \begin{tabular}{l ccccc HcH}
    \toprule
    & 
    \multicolumn{5}{c}{\textbf{nDCG}} &
    \\ 
    \cline{2-6} 
                   \textbf{Model} &  
                   @1  & @2  & @5  & @10  & @20 &
    \textsc{Mean} &
    \textsc{Rank} &
                   \\ 
                   \midrule

Random & 
0.207 & 0.206 & 0.253 & 0.311 & 0.457
&& 5
\\

ConfigPop & 
\text{0.366} & \text{0.532} & \text{0.671} & \text{0.691} & \text{0.693} 
&& 4
\\

\midrule
Ours & 
\textbf{0.827} & \textbf{0.827} & \textbf{0.867} & \textbf{0.882} & \textbf{0.897} 
&& 1
\\

Ours (Deep-only) & 
0.804 & 0.807 & 0.851 & 0.866 & 0.887 
&& 2
\\ 

Ours (Wide-only) & 
0.721 & 0.714 & 0.768 & 0.801 & 0.839 
&& 3
\\
    \bottomrule
    \end{tabular}
\end{table*}

\section{Experiments}
In this section, we conduct experiments to answer the following research questions:

\begin{itemize}
\item \textbf{RQ1:} Given an arbitrary and unseen user selected data set, is our learned model able to automatically recommend the top visualizations that are most important to the user, \ie, the visualizations they manually created, which are held-out for evaluation (Sec.~\ref{sec:exp-quant-results})?

\item \textbf{RQ2:} 
Does our proposed wide-and-deep approach outperform common-sense baselines for end-to-end visualization recommendation?
Is the best performance achieved when using the full wide-and-deep model or do the simpler variants of our approach, namely, using the wide-only or deep-only component of our model perform better (Sec.~\ref{sec:exp-quant-results})?

\item \textbf{RQ3:} Do human experts prefer our ML-based visualization recommendations or the ones from the rule-based system, Voyager2, that uses CompassQL (Sec.~\ref{sec:exp-user-study})?

\item \textbf{RQ4:} Is the learning-based visualization recommendation system able to learn general rules (which would be preferred by human experts) from the large training corpus of datasets and user-generated visualizations (Sec.~\ref{sec:exp-case-study})?
\end{itemize}

To answer RQ1 and RQ2, we quantitatively evaluate our ML-based visualization recommendation system in Section~\ref{sec:exp-quant-results}.
For RQ3, we perform a user study in Section~\ref{sec:exp-user-study} comparing the effectiveness of our ML-based visualization recommender system to the state-of-the-art rule-based system called CompassQL, which is used in both Voyager and Voyager2.
Finally, in Section~\ref{sec:exp-case-study}, we show a number of examples that demonstrate the ability of our ML-based approach to learn rules from the training corpus, and in many cases, perform better than CompassQL (used in Voyager2), and therefore, overcome many of the limitations that exist in such rule-based systems.
These examples also show that our model does not require any manual effort to define such rules, but can automatically learn them from the visualization corpus used for training our model.

\begin{figure*}[t!]
\centering
\subfigure[Top 5 Rule-based Vis Rec.]{
\includegraphics[width=1.0\linewidth]{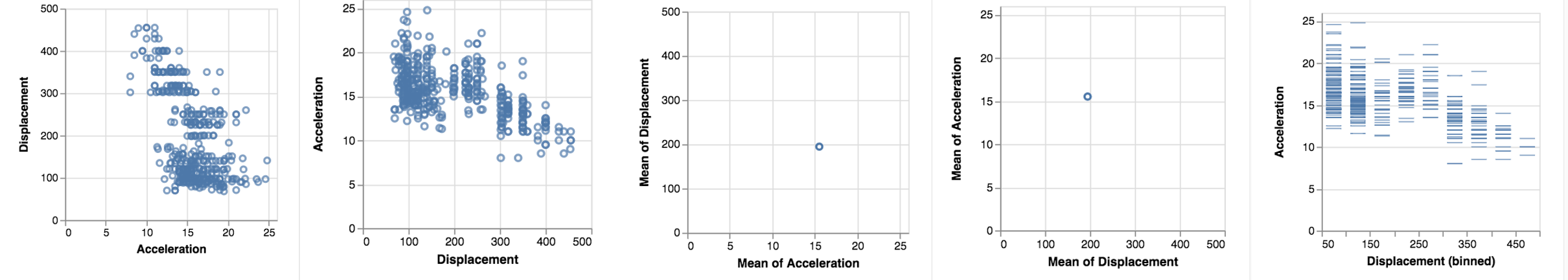}
}
\subfigure[Top 5 ML-based Vis. Rec. (Ours)]{
\includegraphics[width=1.0\linewidth]{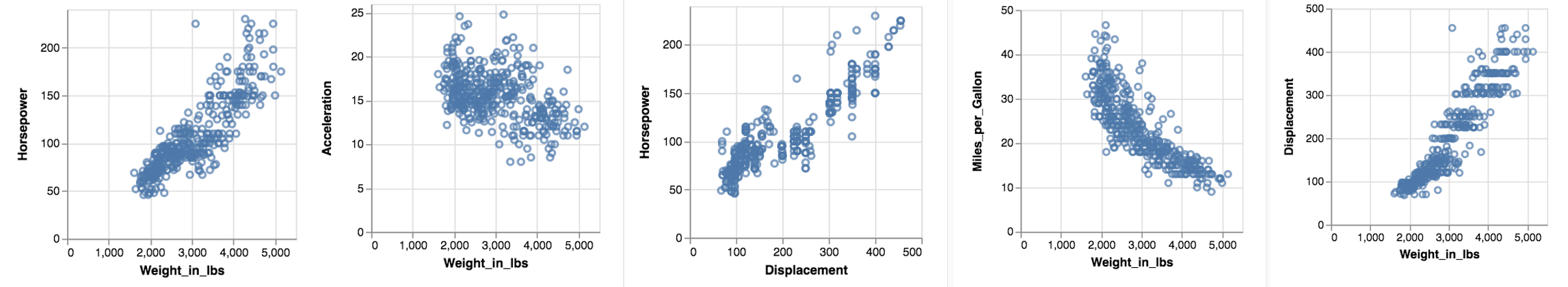}
}
\vspace{-1mm}
\caption{Comparing the top-5 visualization recommendations from the existing end-to-end rule-based system (Voyager2 using CompassQL) to our end-to-end ML-based visualization recommendation system.
See text for discussion.
}
\label{fig-top-5-ML-based-vs-rule-based}
\end{figure*}

\begin{figure*}[t!]
\centering
\includegraphics[width=0.8\linewidth]{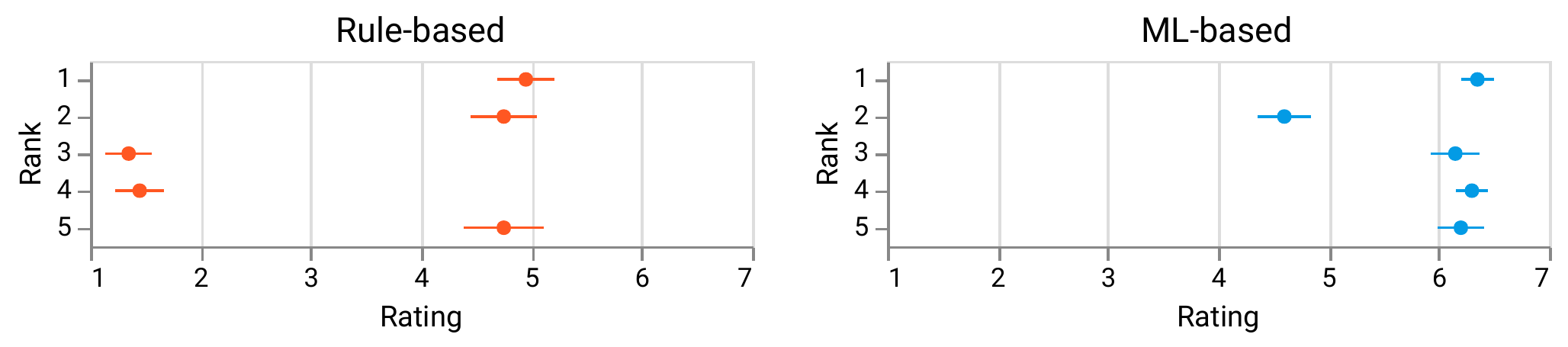}
\vspace{-1mm}
\caption{Human experts' ratings on top 5 visualizations from rule-based (Voyager2 using CompassQL) and our ML-based systems. 
Most visualizations from the ML-based system received higher ratings than rule-base system. 
Strikingly, the top-1 ranked visualization from human experts exactly matches the top-1 visualization recommended by our ML-based model.
Furthermore, the top-4 visualizations receiving the highest rating by human experts are those from the ML-based system.
See the text for discussion on other important findings.
}
\label{fig-top5-mean-scores}
\end{figure*}

\subsection{Quantitative Results}
\label{sec:exp-quant-results}
In this section, we answer RQ1 and RQ2 by evaluating our ML-based visualization recommendation system quantitatively.
For quantitative evaluation, we use the evaluation framework proposed in Section~\ref{sec:eval-framework-MLVisRec}.
For training our ML-based models, we use a training corpus of 1K datasets (and their visualizations) from the \textit{Plot.ly} corpus~\cite{vizml}.
We provide the statistics of the training corpus used for learning our model in Table~\ref{tab:dataset-stats}.
Notably, as shown in Table~\ref{tab:dataset-stats}, there are roughly 1K datasets that have an average of about 12 attributes each.
Our ML-based model for visualization recommendation is learned using 1K datasets consisting of about 12K attributes and about 5K user-generated visualizations that use some of the 12K attributes.

Since this work proposes the first end-to-end learning-based visualization recommender system, we compare our approach using two common-sense baselines along with two simpler variants of our model.
The first common-sense baseline is called random, and refers to a method that 
simply recommends the top-k visualizations chosen uniformly at random from the set of generated visualizations for the specific dataset.
This baseline is important for understanding if our wide-and-deep learning approach is able to learn something meaningful from the raw data (\eg, what visualization preferences, like chart types and so forth users prefer for data with certain characteristics, or what makes a visualization better than another one, and so forth) and if the model is meaningful and useful for visualization recommendation or if it performs no better than random. 
We also propose another common-sense baseline for evaluating ML-based visualization recommender systems based solely on the popularity (or frequency) of a visualization configuration in the training corpus.
We call this baseline ConfigPop.
As an aside, the notion of a visualization configuration, which is proposed in this work, is essentially an abstraction of a visualization, consisting of all the design choices, but not the actual data (or attribute names) used in the visualization, see Section~\ref{sec-problem-formulation} for a more formal definition of the proposed notion of a visualization configuration.

To evaluate the effectiveness of the top recommended visualizations, we use the modified normalized Discounted Cumulative Gain (nDCG)
at $k\in\{1,2,5,10,20\}$ as in Sec.~\ref{sec-eval-metric} for the different top-$k$ visualization recommendations (nDCG@$k$).
Results are reported in Table~\ref{tab:rank-one-seen-among-unseen}.
Strikingly, our wide-and-deep learning-based model for visualization recommendation performs the best, achieving a high nDCG across all $k=1,...,20$, as shown in Table~\ref{tab:rank-one-seen-among-unseen}.
Hence, this confirms that our ML-based visualization recommendation model accurately learns to recommend the top visualizations that are most important to the user, despite that the model has never seen the dataset nor the visualizations created by that user before (RQ1).
In addition, we observe that our full wide-and-deep learning-based approach and the simpler model variants of our approach always outperform the other methods across all $k\in\{1,2,5,10,20\}$.
Furthermore, our approach with both the wide and deep components, has the highest nDCG scores compared to the two common-sense baselines, and our two ablation model variants that use only the wide or deep components of our model, as shown in Table~\ref{tab:rank-one-seen-among-unseen} (RQ2).
It is also important to note that results at smaller $k$ are obviously more important, and these are exactly the situations where our models and the variants perform extremely well compared to the others. As an example, at nDCG@1, our wide-and-deep learning visualization recommendation model achieves 0.827 at $k$=1, whereas the best baseline is only able to achieve an nDCG of 0.366.
This is an improvement in nDCG of 124\% over the best baseline at k=1.

This result clearly demonstrates the effectiveness of our end-to-end ML-based visualization recommendation model as it is able to effectively recover the held-out ground-truth visualization that was generated and therefore preferred by an actual user.
Furthermore, the model is also able to distinguish between visualizations that were not preferred by a user, as they receive a lower score from the ML-based recommendation model.
These results and findings confirm that our model learns to recommend high quality visualizations
that a user will likely prefer 
from an arbitrary and unseen user-selected data set.
From Table~\ref{tab:rank-one-seen-among-unseen}, we also observe that while the full wide-and-deep learning model outperforms our other model variants, the second best performing model is our deep only variant, and it achieves better performance than the wide-only variant across all $k$.
Finally, our ML-based visualization recommendation models always outperform the other baselines across all $k$, and thus are the top 3 best performing models followed by the other baselines that lack any machine learning (using essentially rules, \eg, configPop always predicts the most popular visualization configuration for a given data set).

\subsection{User Study} 
\label{sec:exp-user-study}
In this section, we perform a user study to compare our end-to-end ML-based visualization recommender system to the existing end-to-end system that uses rules as opposed to learning.\footnote{The end-to-end rule-based system used in this study is Voyager2, which uses CompassQL.}
For this, we take the top 5 recommended visualizations from the rule-based system (Voyager2 using CompassQL) and the top-5 recommended visualizations from our ML-based system (for the standard car data).
The top-5 from the rule-based and our ML-based end-to-end visualization recommender system is shown in Figure~\ref{fig-top-5-ML-based-vs-rule-based}.
Given the set of 10 visualizations, we randomize the order by taking a uniform random permutation of them, and then display them to the human experts in this order. 
Human experts are asked to assign a score to each visualization in 7-point Likert scale.
Afterwards, we compute the overall score of a visualization by taking the mean of the scores assigned by the experts.
In this study, there were 21 human experts rating the top-5 visualizations from either system using a 7-point Likert scale.

Comparing the top-5 recommended visualizations from either system, human experts gave significantly higher scores to those visualizations that our ML-based system recommended.
Hence, we observed a strong preference by the human experts towards the visualizations recommended by our system as opposed to the rule-based system.
Overall, the human experts assigned a mean score of 5.92 to the top-5 visualizations from the ML-based approach and a mean score of 3.45 to those from the rule-based approach (RQ3).
Hence, we can clearly see that the ML-based visualization recommendations are significantly better than the rule-based approach.
This result is significant at p-val=$0.01$.
We also provide the mean score and the $\pm$variance for each of the top-5 visualizations from either system in Figure~\ref{fig-top5-mean-scores}.
Strikingly, the top ranked visualization by the human experts is exactly the top ranked visualization from our ML-based system.
Furthermore, among the 10 visualizations that human experts scored, the top 4 visualizations with the highest score are those from our ML-based visualization recommendation system, and not from the rule-based system.
As shown in Figure~\ref{fig-top5-mean-scores}, the variance of the ML-based recommendations are nearly always less than the rule-based system. 
This difference is also significant.
We posit that this is due to the discrete nature of the rule-based recommender system that scores visualization in a discrete fashion using manually defined rules, and so even though the visualization may appear to be high quality with respect to the manually defined rules, it is not of high quality with respect to the actual data and insights that visualizations seek to show from the data.

\begin{figure}[t!]
\begin{center}
\includegraphics[width=1.0\linewidth]{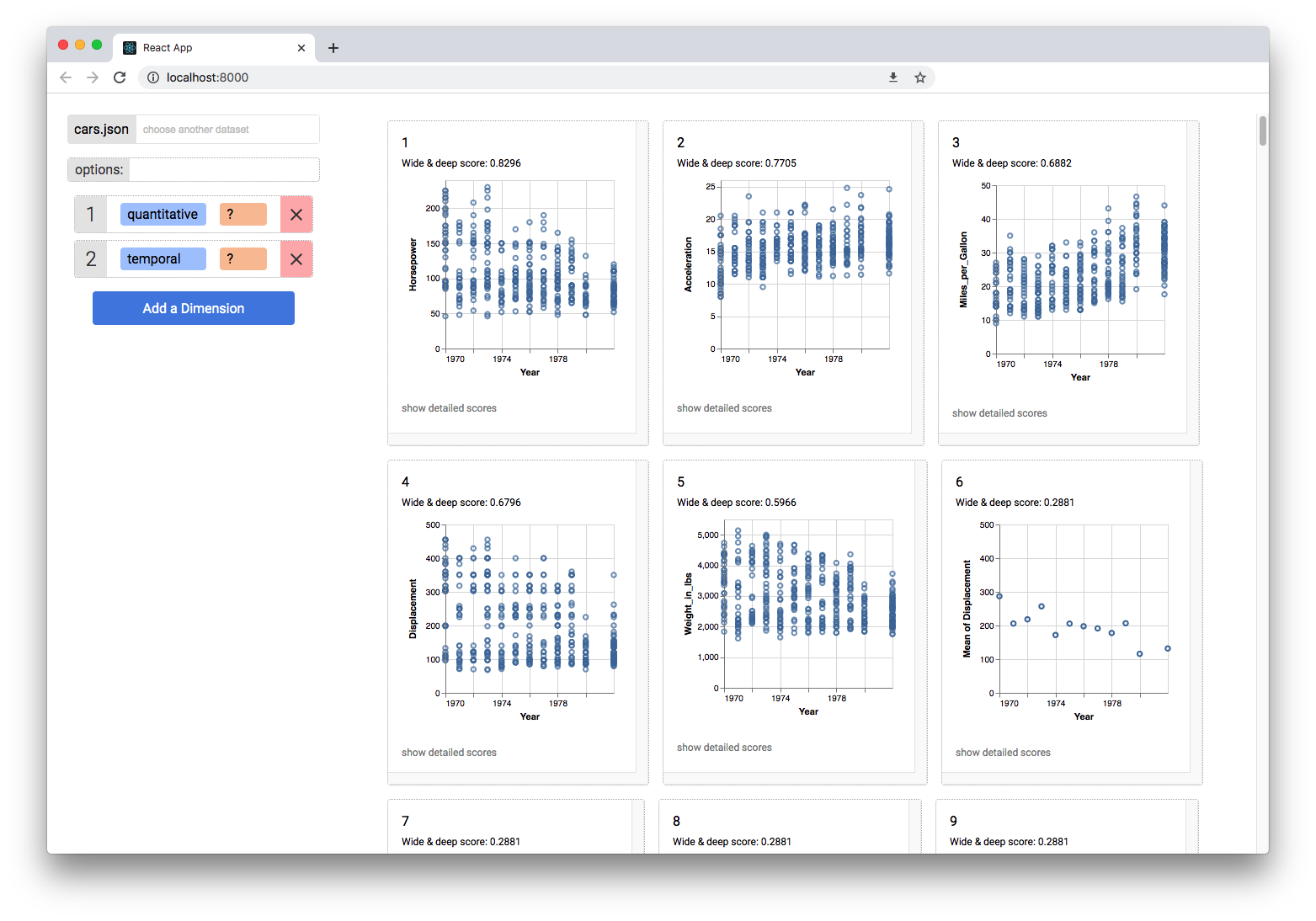}
\end{center}
\caption{
System screenshot of our end-to-end ML-based visualization recommendation approach.
It receives a dataset as input, and shows the top recommended visualizations from our approach ranked by the scores.
} 
\label{fig:anecdotal-interface}
\end{figure}

\subsection{Qualitative Analysis} \label{sec:exp-case-study}
While Section~\ref{sec:exp-quant-results} demonstrated the effectiveness of the visualization ranking from our ML-based approach using a quantitative ranking evaluation metric whereas Section~\ref{sec:exp-user-study} revealed that human experts overwhelmingly preferred visualizations recommended by the ML-based visualization recommendation model compared to those from the state-of-the-art rule-based approach.
In this section, we perform a case study to investigate whether the ML-based visualization recommendation model is able to learn meaningful visual rules from the large training corpus.

To investigate the effectiveness of the visualization recommendations given by our end-to-end ML-based approach, we compare with the existing end-to-end rule-based approach (Voyager2).
As shown in Figure~\ref{fig:anecdotal-interface}, we developed an interface for our ML-based visualization recommendation system that allows the user to select or upload a dataset of interest, and then we automatically recommend them the top visualizations for that given dataset using the learned model.
The recommended visualizations are then displayed to the user in order of relevance/importance score which is inferred from our ML-based model.
In addition, the user can specify different queries interactively using the interface. 
For instance, they can select the attribute types (\textit{quantitative}, \textit{nominal}, and \textit{temporal}) that they want to visualize, and the system immediately infers and displays the top most relevant recommended visualizations to the user.
As an aside, the user can also select attributes of interest to include in the recommended visualizations, aggregations to use, chart-types, and so on.
This is similar to the interface used by Voyager2.

In this case study, we use the \texttt{cars} dataset about car specifications, and specify a few queries.
Results reveal several aspects where our end-to-end ML-based visualization recommendation approach is more effective than the end-to-end rule-based approach (Voyager2), and the ability of our system to automatically learn to recommend visualizations that would be preferred by even domain experts, without the manual specification of any rules (RQ4).

\subsubsection{Learning to place attributes like an expert}
\begin{figure}[]
\centering
\includegraphics[width=1.0\linewidth]{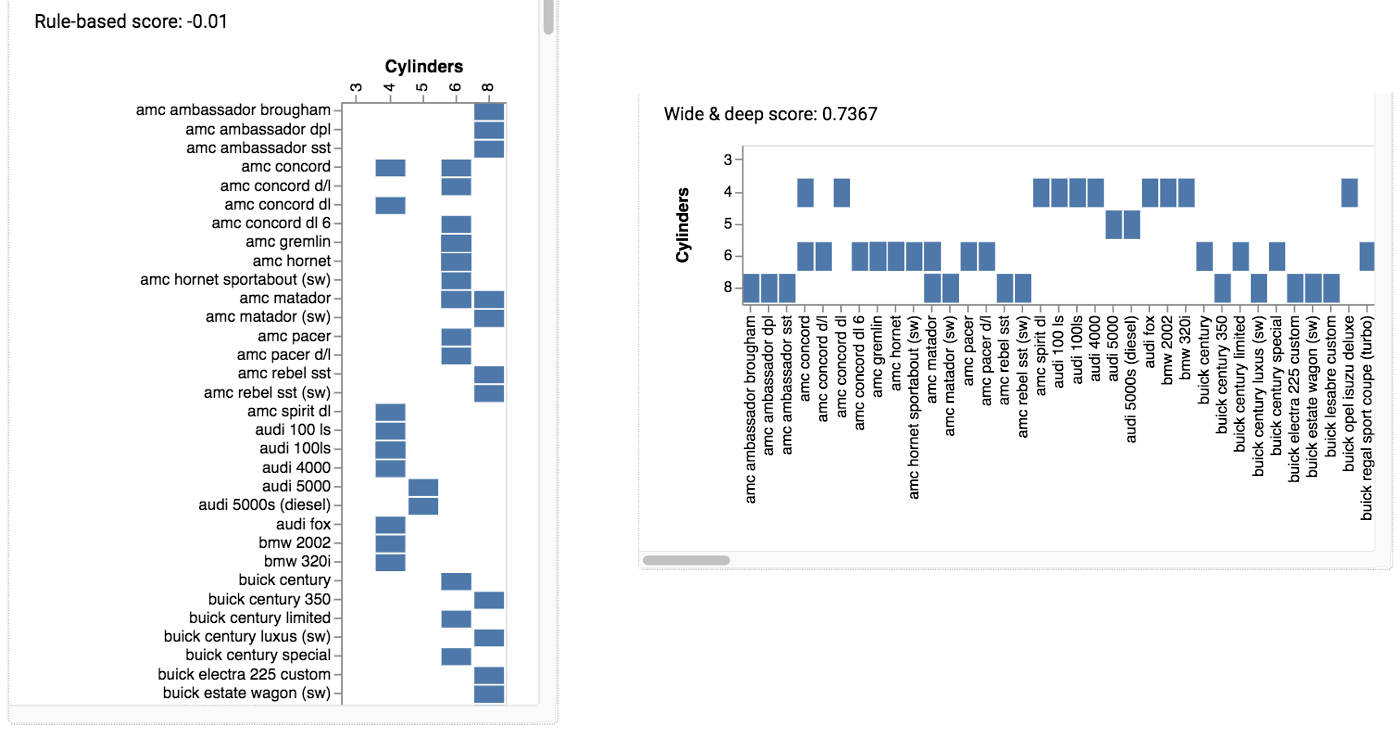}
\caption{
Top visualizations from rule-based (left) vs. our ML-based approach (right) for a query on two nominal attributes.
The ML-based approach learns to recommend a horizontal layout and penalizes the vertical charts.
This is consistent with the fact that domain experts prefer to do it horizontally.
} 
\label{fig-anecdotal-bad-chart-vertical-vs-horizontal-domain-expert-preference}
\end{figure}

Now we demonstrate how our ML-based system is able to learn to prefer visualizations that a human expert would also prefer.
Using a query for visualizations containing two nominal attributes, we see that both approaches recommend visualizations with the attribute car \textit{model name} in their top visualizations as shown in Figure~\ref{fig-anecdotal-bad-chart-vertical-vs-horizontal-domain-expert-preference}.
However, the rule-based system incorrectly recommends a visualization with a vertical layout whereas the ML-based approach learns to recommend a horizontal layout and penalizes the vertical charts.
Interestingly, the ML-based model is able to learn the fact that domain experts prefer to do these types of charts horizontally, as opposed to vertically, and therefore our wide-and-deep learning model penalizes the vertical charts to ensure they are not recommended to the user. This is in complete contrast to the rule-based approach, which seems to favor vertical layout, despite that human experts would recommend against such charts.

\begin{figure}[t!]
\centering
\subfigure[Top 3 Rule-based Vis Rec. (CompassQL/Voyager2)]{
\includegraphics[width=1.0\linewidth]{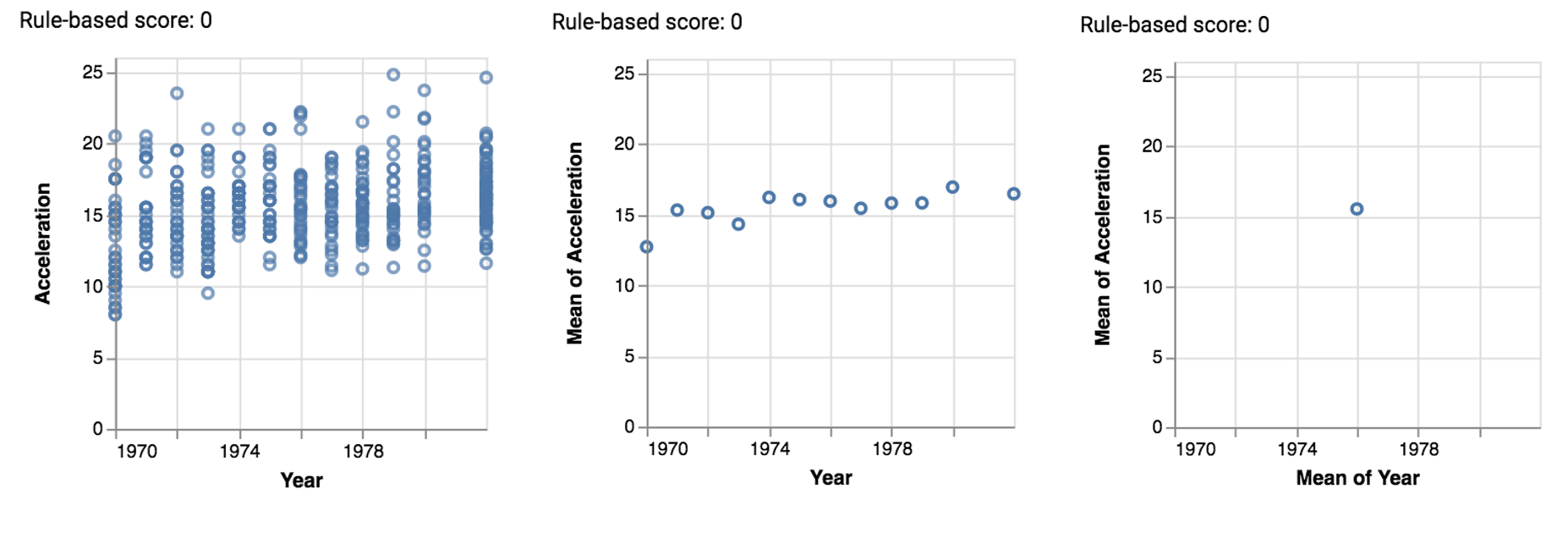}
}
\subfigure[Top 3 ML-based Vis. Rec. (Ours)]{
\includegraphics[width=1.0\linewidth]{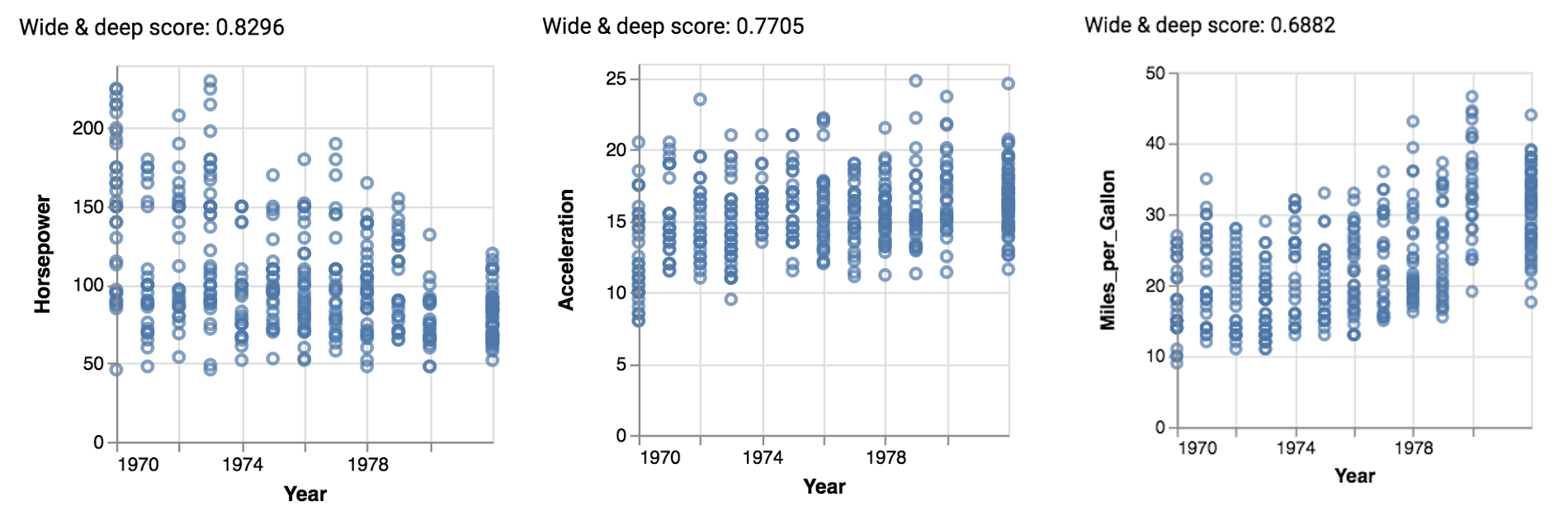}
}
\vspace{-1mm}
\caption{
Tie-breaking issues of the rule-based approach where all visualizations are scored the same and thus the system fails to find an appropriate high quality ranking.
In comparison, our wide-and-deep learning approach is able to learn more meaningful scores that appropriately differentiate between the most important and insightful visualizations and those that are less important.
}
\label{fig-case-study-ML-based-vs-rule-based}
\end{figure}

\subsubsection{Tie-breaking issues}
A key fundamental problem with the existing rule-based visualization recommender system (Voyager2 using CompassQL) is that visualizations are often scored using a set of manually defined rules that assign simple discrete scores to visualizations that either satisfy or violate the manual rules defined by people.
This oversimplified scoring results in many visualizations having exactly the same score, and therefore, no way to actually rank them.
In this case, the existing end-to-end system simply displays the visualizations in the order they were generated, which is most often not very appropriate.
In Figure~\ref{fig-case-study-ML-based-vs-rule-based}, we show one such case of this where the top visualizations from the rule-based are all assigned the same score of 0, and thus fails to prefer one over the other and vice-versa.
Moreover, while we show only three visualizations here with score 0, there are even more further down the list that also have score of 0.
Hence, in this case, the rule-based system simply breaks ties randomly, and often ends up with low quality visualizations that are ranked higher than some visualizations that are clearly better.
In comparison, our wide-and-deep learning approach is able to learn more meaningful scores that appropriately differentiate between the most important and insightful visualizations and those that are less important.
For instance, our ML-based visualization recommendation model assigns a score to the visualization shown in the 3rd position by the rule-based approach, which is clearly not a useful visualization, especially compared to the exponential amount of other possibilities.
There are many other similar situations where our scores are significantly more useful and completely avoid the tie-breaking issues of the rule-based approach.

\section{Conclusion} \label{sec:conc}
In this work, we proposed the first end-to-end ML-based visualization recommendation system.
We first formalized the ML-based visualization recommendation problem and described a generic learning framework for solving it.
Next, we proposed a wide-and-deep learning architecture that combines a wide component with a deep learning component.
Each visualization is scored with the input of two parts, an attribute combination and a visualization configuration, using our wide-and-deep learning visualization recommendation model. 
Given a new unseen dataset from an arbitrary user, the learned model is used to automatically generate, score, and output a list of recommended visualizations for that specific dataset.
Further, we present an evaluation framework that can evaluate visualization recommendation system learned from a large corpus of visualizations and datasets. 
We demonstrated the effectiveness of the proposed approach in three different ways.
First, we used the evaluation framework to quantitatively demonstrate the effectiveness of the ranking of visualizations from our ML-based visualization recommendation system.
Second, we also validated the effectiveness of our ML-based system through a user study of 20 human experts, and found that the top-1 ranked visualization from human experts matched exactly the top-1 visualization from our system.
Most importantly, human experts overwhelming preferred the ML-based visualization recommendations over the existing rule-based system as shown in Section~\ref{sec:exp-user-study}.
Third, we also performed qualitative analysis on the recommendations from our ML-based model and the state-of-the-art rule-based approach, and discussed many different advantages where our model is able to learn visual rules from the large training corpus that are not even incorporated in the rule-based approach.

\bibliographystyle{ACM-Reference-Format}
\bibliography{references}


\begin{thebibliography}{18}


\ifx \showCODEN    \undefined \def \showCODEN     #1{\unskip}     \fi
\ifx \showDOI      \undefined \def \showDOI       #1{#1}\fi
\ifx \showISBNx    \undefined \def \showISBNx     #1{\unskip}     \fi
\ifx \showISBNxiii \undefined \def \showISBNxiii  #1{\unskip}     \fi
\ifx \showISSN     \undefined \def \showISSN      #1{\unskip}     \fi
\ifx \showLCCN     \undefined \def \showLCCN      #1{\unskip}     \fi
\ifx \shownote     \undefined \def \shownote      #1{#1}          \fi
\ifx \showarticletitle \undefined \def \showarticletitle #1{#1}   \fi
\ifx \showURL      \undefined \def \showURL       {\relax}        \fi
\providecommand\bibfield[2]{#2}
\providecommand\bibinfo[2]{#2}
\providecommand\natexlab[1]{#1}
\providecommand\showeprint[2][]{arXiv:#2}

\bibitem[\protect\citeauthoryear{Ehsan, Sharaf, and Chrysanthis}{Ehsan
  et~al\mbox{.}}{2016}]%
        {ehsan2016muve}
\bibfield{author}{\bibinfo{person}{Humaira Ehsan}, \bibinfo{person}{Mohamed~A
  Sharaf}, {and} \bibinfo{person}{Panos~K Chrysanthis}.}
  \bibinfo{year}{2016}\natexlab{}.
\newblock \showarticletitle{Muve: Efficient multi-objective view recommendation
  for visual data exploration}. In \bibinfo{booktitle}{\emph{ICDE '16}}.
  \bibinfo{pages}{731--742}.
\newblock


\bibitem[\protect\citeauthoryear{He and Chua}{He and Chua}{2017}]%
        {he2017neural}
\bibfield{author}{\bibinfo{person}{Xiangnan He} {and} \bibinfo{person}{Tat-Seng
  Chua}.} \bibinfo{year}{2017}\natexlab{}.
\newblock \showarticletitle{Neural factorization machines for sparse predictive
  analytics}. In \bibinfo{booktitle}{\emph{SIGIR '17}}.
  \bibinfo{pages}{355--364}.
\newblock


\bibitem[\protect\citeauthoryear{Hu, Bakker, Li, Kraska, and Hidalgo}{Hu
  et~al\mbox{.}}{2019}]%
        {vizml}
\bibfield{author}{\bibinfo{person}{Kevin Hu}, \bibinfo{person}{Michiel~A
  Bakker}, \bibinfo{person}{Stephen Li}, \bibinfo{person}{Tim Kraska}, {and}
  \bibinfo{person}{C{\'e}sar Hidalgo}.} \bibinfo{year}{2019}\natexlab{}.
\newblock \showarticletitle{Vizml: A machine learning approach to visualization
  recommendation}. In \bibinfo{booktitle}{\emph{CHI '19}}.
  \bibinfo{pages}{1--12}.
\newblock


\bibitem[\protect\citeauthoryear{Hu, Orghian, and Hidalgo}{Hu
  et~al\mbox{.}}{2018}]%
        {hu2018dive}
\bibfield{author}{\bibinfo{person}{Kevin Hu}, \bibinfo{person}{Diana Orghian},
  {and} \bibinfo{person}{C{\'e}sar Hidalgo}.} \bibinfo{year}{2018}\natexlab{}.
\newblock \showarticletitle{Dive: A mixed-initiative system supporting
  integrated data exploration workflows}. In
  \bibinfo{booktitle}{\emph{Proceedings of the Workshop on Human-In-the-Loop
  Data Analytics}}. \bibinfo{pages}{1--7}.
\newblock


\bibitem[\protect\citeauthoryear{Lee}{Lee}{2020}]%
        {dorisjang}
\bibfield{author}{\bibinfo{person}{Doris Jung-Lin Lee}.}
  \bibinfo{year}{2020}\natexlab{}.
\newblock \bibinfo{booktitle}{\emph{Insight Machines: The Past, Present, and
  Future of Visualization Recommendation}}.
\newblock


\bibitem[\protect\citeauthoryear{Lee, Dev, Hu, Elmeleegy, and Parameswaran}{Lee
  et~al\mbox{.}}{2019}]%
        {lee2019avoiding}
\bibfield{author}{\bibinfo{person}{Doris Jung-Lin Lee}, \bibinfo{person}{Himel
  Dev}, \bibinfo{person}{Huizi Hu}, \bibinfo{person}{Hazem Elmeleegy}, {and}
  \bibinfo{person}{Aditya Parameswaran}.} \bibinfo{year}{2019}\natexlab{}.
\newblock \showarticletitle{Avoiding drill-down fallacies with VisPilot:
  assisted exploration of data subsets}. In \bibinfo{booktitle}{\emph{IUI
  '19}}. \bibinfo{pages}{186--196}.
\newblock


\bibitem[\protect\citeauthoryear{Mackinlay}{Mackinlay}{1986}]%
        {mackinlay1986automating}
\bibfield{author}{\bibinfo{person}{Jock Mackinlay}.}
  \bibinfo{year}{1986}\natexlab{}.
\newblock \showarticletitle{Automating the design of graphical presentations of
  relational information}.
\newblock \bibinfo{journal}{\emph{ACM Trans. Graph.}} \bibinfo{volume}{5},
  \bibinfo{number}{2} (\bibinfo{year}{1986}), \bibinfo{pages}{110--141}.
\newblock


\bibitem[\protect\citeauthoryear{Mackinlay, Hanrahan, and Stolte}{Mackinlay
  et~al\mbox{.}}{2007}]%
        {mackinlay2007show}
\bibfield{author}{\bibinfo{person}{Jock Mackinlay}, \bibinfo{person}{Pat
  Hanrahan}, {and} \bibinfo{person}{Chris Stolte}.}
  \bibinfo{year}{2007}\natexlab{}.
\newblock \showarticletitle{Show me: Automatic presentation for visual
  analysis}.
\newblock \bibinfo{journal}{\emph{IEEE transactions on visualization and
  computer graphics}} \bibinfo{volume}{13}, \bibinfo{number}{6}
  (\bibinfo{year}{2007}), \bibinfo{pages}{1137--1144}.
\newblock


\bibitem[\protect\citeauthoryear{Moritz, Wang, Nelson, Lin, Smith, Howe, and
  Heer}{Moritz et~al\mbox{.}}{2018}]%
        {draco}
\bibfield{author}{\bibinfo{person}{Dominik Moritz}, \bibinfo{person}{Chenglong
  Wang}, \bibinfo{person}{Greg~L Nelson}, \bibinfo{person}{Halden Lin},
  \bibinfo{person}{Adam~M Smith}, \bibinfo{person}{Bill Howe}, {and}
  \bibinfo{person}{Jeffrey Heer}.} \bibinfo{year}{2018}\natexlab{}.
\newblock \showarticletitle{Formalizing visualization design knowledge as
  constraints: Actionable and extensible models in draco}.
\newblock \bibinfo{journal}{\emph{IEEE Transactions on Visualization and
  Computer Graphics}} \bibinfo{volume}{25}, \bibinfo{number}{1}
  (\bibinfo{year}{2018}), \bibinfo{pages}{438--448}.
\newblock


\bibitem[\protect\citeauthoryear{Perry, Howe, Key, and Aragon}{Perry
  et~al\mbox{.}}{2013}]%
        {perry2013vizdeck}
\bibfield{author}{\bibinfo{person}{Daniel~B Perry}, \bibinfo{person}{Bill
  Howe}, \bibinfo{person}{Alicia~MF Key}, {and} \bibinfo{person}{Cecilia
  Aragon}.} \bibinfo{year}{2013}\natexlab{}.
\newblock \showarticletitle{VizDeck: Streamlining exploratory visual analytics
  of scientific data}. In \bibinfo{booktitle}{\emph{iConference '13}}.
\newblock


\bibitem[\protect\citeauthoryear{Roth, Kolojejchick, Mattis, and
  Goldstein}{Roth et~al\mbox{.}}{1994}]%
        {roth1994interactive}
\bibfield{author}{\bibinfo{person}{Steven~F Roth}, \bibinfo{person}{John
  Kolojejchick}, \bibinfo{person}{Joe Mattis}, {and} \bibinfo{person}{Jade
  Goldstein}.} \bibinfo{year}{1994}\natexlab{}.
\newblock \showarticletitle{Interactive graphic design using automatic
  presentation knowledge}. In \bibinfo{booktitle}{\emph{CHI '94}}.
  \bibinfo{pages}{112--117}.
\newblock


\bibitem[\protect\citeauthoryear{Sarawagi, Agrawal, and Megiddo}{Sarawagi
  et~al\mbox{.}}{1998}]%
        {sarawagi1998discovery}
\bibfield{author}{\bibinfo{person}{Sunita Sarawagi}, \bibinfo{person}{Rakesh
  Agrawal}, {and} \bibinfo{person}{Nimrod Megiddo}.}
  \bibinfo{year}{1998}\natexlab{}.
\newblock \showarticletitle{Discovery-driven exploration of OLAP data cubes}.
  In \bibinfo{booktitle}{\emph{International Conference on Extending Database
  Technology}}. Springer, \bibinfo{pages}{168--182}.
\newblock


\bibitem[\protect\citeauthoryear{Vartak, Huang, Siddiqui, Madden, and
  Parameswaran}{Vartak et~al\mbox{.}}{2017}]%
        {vartak2017towards}
\bibfield{author}{\bibinfo{person}{Manasi Vartak}, \bibinfo{person}{Silu
  Huang}, \bibinfo{person}{Tarique Siddiqui}, \bibinfo{person}{Samuel Madden},
  {and} \bibinfo{person}{Aditya Parameswaran}.}
  \bibinfo{year}{2017}\natexlab{}.
\newblock \showarticletitle{Towards visualization recommendation systems}.
\newblock \bibinfo{journal}{\emph{ACM SIGMOD Record}} \bibinfo{volume}{45},
  \bibinfo{number}{4} (\bibinfo{year}{2017}), \bibinfo{pages}{34--39}.
\newblock


\bibitem[\protect\citeauthoryear{Wilkinson, Anand, and Grossman}{Wilkinson
  et~al\mbox{.}}{2005}]%
        {wilkinson2005graph}
\bibfield{author}{\bibinfo{person}{Leland Wilkinson}, \bibinfo{person}{Anushka
  Anand}, {and} \bibinfo{person}{Robert Grossman}.}
  \bibinfo{year}{2005}\natexlab{}.
\newblock \showarticletitle{Graph-theoretic scagnostics}. In
  \bibinfo{booktitle}{\emph{IEEE Symposium on Information Visualization}}.
  IEEE, \bibinfo{pages}{157--164}.
\newblock


\bibitem[\protect\citeauthoryear{Wills and Wilkinson}{Wills and
  Wilkinson}{2010}]%
        {wills2010autovis}
\bibfield{author}{\bibinfo{person}{Graham Wills} {and} \bibinfo{person}{Leland
  Wilkinson}.} \bibinfo{year}{2010}\natexlab{}.
\newblock \showarticletitle{Autovis: automatic visualization}.
\newblock \bibinfo{journal}{\emph{Information Visualization}}
  \bibinfo{volume}{9}, \bibinfo{number}{1} (\bibinfo{year}{2010}),
  \bibinfo{pages}{47--69}.
\newblock


\bibitem[\protect\citeauthoryear{Wongsuphasawat, Moritz, Anand, Mackinlay,
  Howe, and Heer}{Wongsuphasawat et~al\mbox{.}}{2015}]%
        {wongsuphasawat2015voyager}
\bibfield{author}{\bibinfo{person}{Kanit Wongsuphasawat},
  \bibinfo{person}{Dominik Moritz}, \bibinfo{person}{Anushka Anand},
  \bibinfo{person}{Jock Mackinlay}, \bibinfo{person}{Bill Howe}, {and}
  \bibinfo{person}{Jeffrey Heer}.} \bibinfo{year}{2015}\natexlab{}.
\newblock \showarticletitle{Voyager: Exploratory analysis via faceted browsing
  of visualization recommendations}.
\newblock \bibinfo{journal}{\emph{IEEE transactions on visualization and
  computer graphics}} \bibinfo{volume}{22}, \bibinfo{number}{1}
  (\bibinfo{year}{2015}), \bibinfo{pages}{649--658}.
\newblock


\bibitem[\protect\citeauthoryear{Wongsuphasawat, Moritz, Anand, Mackinlay,
  Howe, and Heer}{Wongsuphasawat et~al\mbox{.}}{2016}]%
        {wongsuphasawat2016towards}
\bibfield{author}{\bibinfo{person}{Kanit Wongsuphasawat},
  \bibinfo{person}{Dominik Moritz}, \bibinfo{person}{Anushka Anand},
  \bibinfo{person}{Jock Mackinlay}, \bibinfo{person}{Bill Howe}, {and}
  \bibinfo{person}{Jeffrey Heer}.} \bibinfo{year}{2016}\natexlab{}.
\newblock \showarticletitle{Towards a general-purpose query language for
  visualization recommendation}. In \bibinfo{booktitle}{\emph{Proceedings of
  the Workshop on Human-In-the-Loop Data Analytics}}. \bibinfo{pages}{1--6}.
\newblock


\bibitem[\protect\citeauthoryear{Wongsuphasawat, Qu, Moritz, Chang, Ouk, Anand,
  Mackinlay, Howe, and Heer}{Wongsuphasawat et~al\mbox{.}}{2017}]%
        {wongsuphasawat2017voyager}
\bibfield{author}{\bibinfo{person}{Kanit Wongsuphasawat},
  \bibinfo{person}{Zening Qu}, \bibinfo{person}{Dominik Moritz},
  \bibinfo{person}{Riley Chang}, \bibinfo{person}{Felix Ouk},
  \bibinfo{person}{Anushka Anand}, \bibinfo{person}{Jock Mackinlay},
  \bibinfo{person}{Bill Howe}, {and} \bibinfo{person}{Jeffrey Heer}.}
  \bibinfo{year}{2017}\natexlab{}.
\newblock \showarticletitle{Voyager 2: Augmenting visual analysis with partial
  view specifications}. In \bibinfo{booktitle}{\emph{CHI '17}}.
  \bibinfo{pages}{2648--2659}.
\newblock


\end{thebibliography}

\end{document}